 \documentclass[reprint,twocolumn,superscriptaddress,secnumarabic,amssymb, nobibnotes, aps,pra, 
 ]{revtex4-2}

\usepackage{graphicx}%
\usepackage{dcolumn}
\usepackage{amsmath}
\usepackage{amssymb}
\usepackage{bm}
\usepackage{color}
\usepackage[whole]{bxcjkjatype}
\usepackage{xspace}
\usepackage{comment}

\begin{document}



\preprint{submitted to Physical Review Letters}

\title{Fingerprints of Mott and Slater gaps in the core-level photoemission spectra of antiferromagnetic iridates}

\author{K. Nakagawa}
\affiliation{Graduate School of Natural Science, Konan University, Kobe 658-8501, Japan }
\affiliation{RIKEN SPring-8 Center, Sayo, Hyogo 679-5148, Japan}

\author{A. Hariki}
\affiliation{Graduate School of Engineering, Osaka Prefecture University, Sakai, Osaka 599-8531, Japan}
\affiliation{Graduate School of Engineering, Osaka Metropolitan University, Sakai, Osaka 599-8531, Japan}

\author{T. Okauchi}
\affiliation{Graduate School of Engineering, Osaka Metropolitan University, Sakai, Osaka 599-8531, Japan}

\author{H. Fujiwara}
\affiliation{RIKEN SPring-8 Center, Sayo, Hyogo 679-5148, Japan}
\affiliation{Graduate School of Engineering Science, The University of Osaka, Toyonaka, Osaka 560-8531, Japan}
\affiliation{
Spintronics Research Network Division, 
Institute for Open and Transdisciplinary Research Initiatives,
The University of Osaka, 
Suita, Osaka 565-0871, Japan}

\author{K.-H. Ahn}   
\affiliation{Institute of Physics, Czech Academy of Sciences, Cukrovarnick\'{a} 10, 162 00 Praha 6, Czechia}

\author{Y. Murakami}
\affiliation{Graduate School of Natural Science, Konan University, Kobe 658-8501, Japan }
\affiliation{RIKEN SPring-8 Center, Sayo, Hyogo 679-5148, Japan}

\author{S.~Hamamoto}
\affiliation{RIKEN SPring-8 Center, Sayo, Hyogo 679-5148, Japan}

\author{Y. Kanai-Nakata}
\author{T. Kadono}
\affiliation{RIKEN SPring-8 Center, Sayo, Hyogo 679-5148, Japan}
\affiliation{College of Science and Engineering, Ritsumeikan University, Kusatsu, Shiga 525-8577, Japan}

\author{A. Higashiya}
\affiliation{RIKEN SPring-8 Center, Sayo, Hyogo 679-5148, Japan}
\affiliation{Faculty of Science and Engineering, Setsunan University, Neyagawa, Osaka 572-8508, Japan}

\author{K. Tamasaku}
\author{M.~Yabashi}
\author{T. Ishikawa}
\affiliation{RIKEN SPring-8 Center, Sayo, Hyogo 679-5148, Japan}

\author{A. Sekiyama}
\affiliation{RIKEN SPring-8 Center, Sayo, Hyogo 679-5148, Japan}
\affiliation{Graduate School of Engineering Science, The University of Osaka, Toyonaka, Osaka 560-8531, Japan}
\affiliation{
Spintronics Research Network Division, 
Institute for Open and Transdisciplinary Research Initiatives,
The University of Osaka, Suita, Osaka 565-0871, Japan}

\author{S. Imada}
\affiliation{RIKEN SPring-8 Center, Sayo, Hyogo 679-5148, Japan}
\affiliation{College of Science and Engineering, Ritsumeikan University, Kusatsu, Shiga 525-8577, Japan}

\author{J. Kune\v{s}}
\affiliation{Institute of Solid State Physics, TU Wien, 1040 Vienna, Austria}
\affiliation{Department of Condensed Matter Physics, Faculty of
  Science, Masaryk University, Kotl\'a\v{r}sk\'a 2, 611 37 Brno,
  Czechia}

\author{K. Takase}
\affiliation{College of Science and Technology, Nihon University, Tokyo 101-8308, Japan}

\author{A. Yamasaki}
\email[Author to whom correspondence should be addressed. \\
E-mail: ]{yamasaki@konan-u.ac.jp}
\affiliation{RIKEN SPring-8 Center, Sayo, Hyogo 679-5148, Japan}
\affiliation{Faculty of Science and Engineering, Konan University, Kobe 658-8501, Japan }

\date{\today}

\begin{abstract}

We present Ir $4f$ core-level hard-x-ray photoemission spectroscopy (HAXPES) experiments conducted across the antiferromagnetic (AFM) ordering transition in Ruddlesden-Popper iridates Sr$_{2}$IrO$_{4}$ and Sr$_{3}$Ir$_{2}$O$_{7}$. The Ir $4f$ spectra exhibit distinct changes between the AFM and paramagnetic (PM) phases, with the spectral difference $I_\text{PM}-I_\text{AFM}$ showing a contrasting behavior in the two compounds.  By employing computational simulations using the local density approximation  combined with the dynamical mean-field theory  method, we elucidate that $I_\text{PM}-I_\text{AFM}$ primarily reflects the Slater or Mott-Hubbard character of the AFM insulating state rather than material-specific details, such as crystal and/or band structures. This sensitivity to fine low-energy electronic structure arises from the dependence of charge-transfer responses to 
the sudden creation of a localized core hole on both metal-insulator transitions and long-range AFM ordering. Our result broadens the applications of core-level HAXPES as a tool for characterization of electronic structure in 5$d$ transition-metal compounds.

\end{abstract}



\maketitle

%
%
\section{Introduction}
Antiferromagnetic (AFM) materials often exhibit gaps in their charge excitation spectra. Their origin can be traced either to strong electronic correlations 
present 
also 
in the paramagnetic (PM) phase, the Mott mechanism, or to the long-range AFM order and associated doubling of the unit cell, the Slater mechanism. Naively, the distinction between the two mechanisms is straightforward: ``Is there an intrinsic charge gap in the PM phase or not?'' The reality and its experimental determination are more complex. Detecting a gap using angle-resolved photoemission spectroscopy (ARPES) may prove difficult~\cite{King13,Yamasaki16}. Besides an instrumental resolution limitation, at higher N\'eel transition temperatures $T_N$ thermal broadening or possible coupling with bosonic excitations obscures small gaps. Moreover, changes in the ARPES
spectra are very specific to
the 
band structure of a given material, which makes identification 
of the Mott or Slater characteristics complicated. 

Core-level photoemission spectroscopy, which measures the time evolution of the core hole left by the photoelectron, does not suffer from the above problems
since it is a local probe. The localized core hole represents strong perturbation to the electronic system and triggers dynamical charge response of surrounding valence electrons, traditionally called charge-transfer screening, providing a sensitivity to the low-energy states near the Fermi energy ($E_F$).  On the other hand, the ability of a local probe
to provide  useful information about long-range AFM order is not obvious. 

To answer the Mott vs Slater question mentioned above, Ruddlesden-Popper iridates are suitable as test materials. They provide an excellent opportunity to study the phase transition triggered by both the electron correlation and the magnetic interaction, since the strength of the electron correlation can be tuned by the dimensionality of the IrO$_2$ plane structure.  Single-layer Sr$_{2}$IrO$_{4}$  is widely known as a spin-orbit assisted Mott insulator with $T_N\simeq$240~K~\cite{Cao1998}, while the double-layered AFM insulator Sr$_{3}$Ir$_{2}$O$_{7}$ ($T_N\simeq$280~K)~\cite{Fujiyama12} is thought to be close to the metal-insulator transition.

In this article, we show that it is possible to address the question using the {\it core-level} hard-x-ray photoemission spectroscopy (HAXPES). 
We observe a clear difference in the behavior of Ir 4$f$ core-level HAXPES spectra at the AFM transition between these materials.
The experimental spectra are well reproduced by theoretical calculations based on the local density approximation (LDA) + dynamical mean-field theory (DMFT), which provide an explanation for the observed behavior and link it to a dominant Mott or Slater mechanism.

\section{Experiment}
Single crystals of Sr$_{2}$IrO$_{4}$ and  Sr$_{3}$Ir$_{2}$O$_{7}$ were grown by the flux method using SrCl$_2$ as the flux material; see Ref.~\onlinecite{Yamasaki16} for the sample preparation and their characterization.
The normal-emission HAXPES data were collected at the beamline BL19LXU of SPring-8~\cite{Yabashi_01}. 
The linearly polarized x-ray delivered by the 27m-long undulator was monochromized by the Si (111) double-crystal and Si (620) channel-cut monochromators~\cite{Fujiwara16}. 
The angle between the $p$-polarized x-ray and the entrance of the photoelectron analyzer was set to 60 degrees in the plane of incidence.
The photon energy was chosen to be 7.9~keV with the total energy resolution of $\Delta E=400$~meV as the full width at half maximum (FWHM). 
The (001) surfaces were obtained by cleaving the samples {\it in situ} in ultrahigh vacuum ($\leq 1 \times10^{-7}$ Pa).

No contamination was detected in the survey spectra.

\section{Theory}
The LDA+DMFT calculations  are performed for the experimental crystal structures of Sr$_{2}$IrO$_{4}$ and Sr$_{3}$Ir$_{2}$O$_{7}$ with the implementation used in Refs.~\onlinecite{Hariki17,Rahn22,Higashi2021}. The $dp$ tight-binding model spanning the Ir $5d$ and O 2$p$ bands derived from the LDA bands is augmented by the local electron-electron interaction on the Ir 5$d$ shell parametrized by Hubbard $U=4.5$~eV and Hund's $J=0.8$~eV.
These values are consistent with previous 
studies 
based on the density functional theory (DFT) + DMFT
for iridates~\cite{Zhang2013}. 
The double-counting correction $\mu_{\rm dc}$ is introduced in order to remove the interaction effects already present in the LDA description~\cite{Kotliar06,Karolak10}. The bare Ir $5d$ site energy $\varepsilon_d^{\rm LDA}$ is shifted by the $\mu_{\rm dc}$, which modifies the Ir $5d$--O 2$p$ splitting and, consequently, the Ir 5$d$ band-width and the metal-insulator transition (MIT)~\cite{Higashi2021}. 
This allows us to use $\mu_{\rm dc}$ as a tuning parameter
to go between Slater and Mott-Hubbard regimes in the same material
and thus demonstrate the impact on the core-level spectra. 
The realistic values of  $\mu_{\rm dc}$ are obtained by comparison to the experimental core-level and valence-band HAXPES spectra. 
The continuous-time hybridization expansion Monte Carlo method~\cite{Werner2006a} is employed as the DMFT impurity solver.
In simulating the AFM phase, we allow for spin dependence in the self-energy $\Sigma(\omega)$ in the DMFT self-consistent calculation. We found that the experimental AFM structures~\cite{Kim12,Kim09} are stable in both compounds.
The spectral functions and hybridization densities 
on the real-frequency axis are computed with the analytically continued  $\Sigma(\omega)$ by the maximum entropy method~\cite{jarrell96}. 
The
Ir core-level photoemission spectra are simulated
using the Anderson impurity model (AIM) with the DMFT hybridization densities
where the core-valence interaction (core-hole potential) $U_{cd}$ is included explicitly in the 
photoemission final states~\cite{Hariki17,Ghiasi19}.  
As we explain details below and in the 
Appendix B,
we neglect core-valence multiplet effects in the AIM.

\section{results and discussion}
\begin{figure}[tb]
\includegraphics[width=0.9\columnwidth]{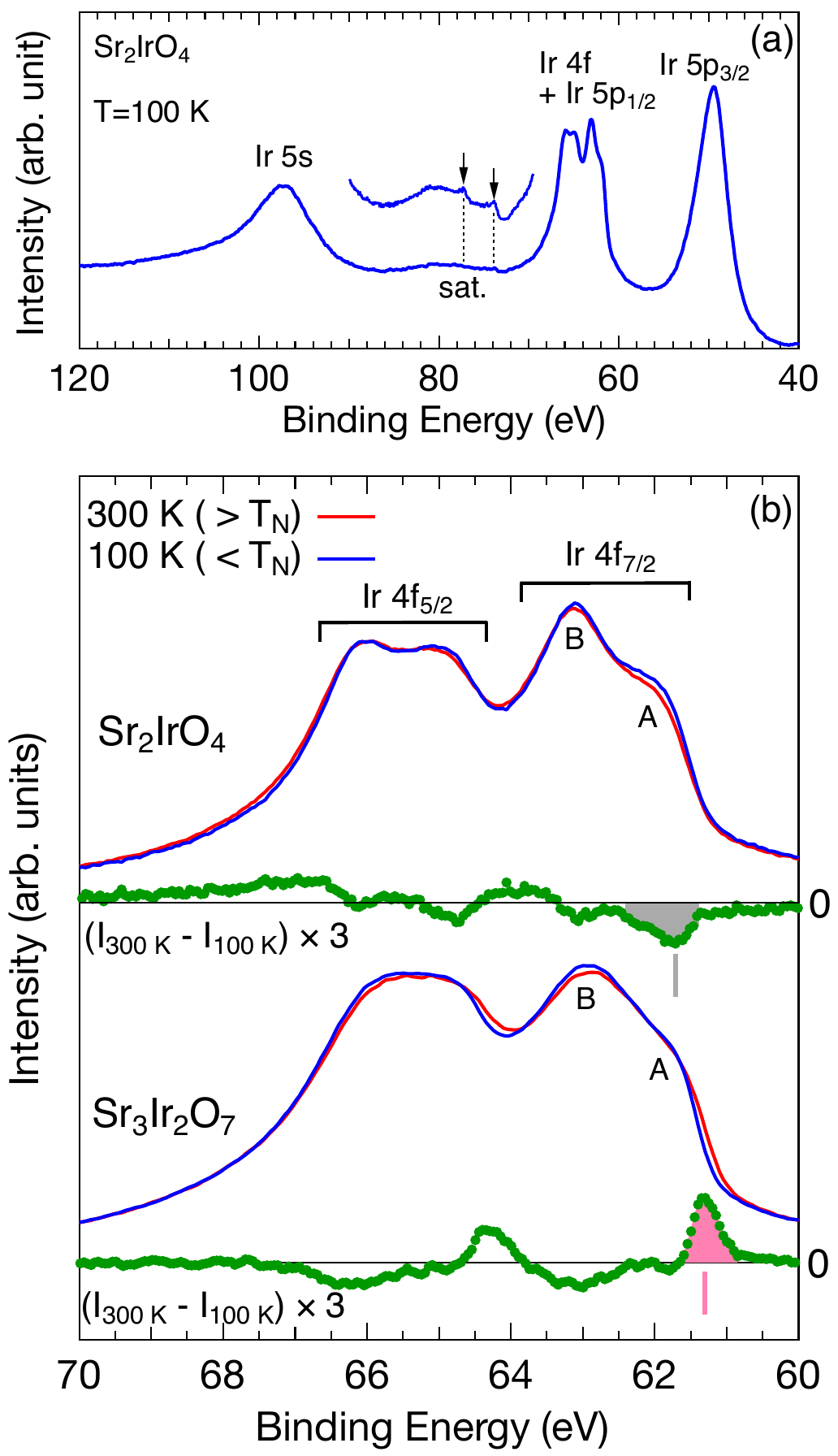}
\caption{(a) HAXPES spectrum of Ir core levels in Sr$_{2}$IrO$_{4}$. (b) Temperature dependence of Ir $4f$  spectra for Sr$_{2}$IrO$_{4}$ and Sr$_{3}$Ir$_{2}$O$_{7}$. The spectra are normalized by the area under the curves after subtraction of the Shirley-type background~\cite{Shirley72}.
The difference spectra are also shown.
}
\label{fig_exp}
\end{figure}
Figure~\ref{fig_exp}(a) shows the core-level HAXPES spectrum of Sr$_{2}$IrO$_{4}$ in a wide binding energy ($E_B$) range including the Ir $4f$ core levels.
Quasisymmetric peaks corresponding to Ir $5s$  and  $5p_{3/2}$ states are found
 around 100~eV and 50~eV, respectively.
A complex spectral feature 
can be seen between them,
which is attributed to Ir $4f$ and $5p_{1/2}$ states.
According to previous studies of metallic Ir and IrO$_2$~\cite{Freakley16}, the Ir $5p_{1/2}$ state has the same $E_B$ as the Ir $4f$ states.
Some spectral weights due to a Sr $4s$ plasmon satellite ($E_B\simeq66$~eV) also contribute
to form the complex Ir $4f$ spectral shape in strontium iridates.
The $5p_{1/2}$ and plasmon contributions on the $4f$ states 
are quantitatively evaluated by a line-shape analysis shown in 
the Supplemental Material (SM)~\cite{sm}.
Unlike the $4f$ features,
the broader $5p_{1/2}$ peak and plasmon satellite are essentially insensitive to the magnetic order and thus
their contribution cancels out in the spectral difference $I_\text{300 K}-I_\text{100 K}$~\footnote{The nonlocal screening effect, discussed in the Ir 4$f$ spectra below, could also be present in the Ir 5$p_{1/2}$ core-level spectra as well. However, the Ir 5$p_{1/2}$ line width is approximately six times broader compared to the Ir 4$f$ core level due to large lifetime broadening~\cite{sm}. Thus, the temperature dependence should be negligibly weak in the Ir 5$p_{1/2}$ spectra. 
}.
In both Sr$_{2}$IrO$_{4}$ and Sr$_{3}$Ir$_{2}$O$_{7}$, fine features labeled as $A$ and $B$ in the Ir $4f_{7/2}$ component can be identified; 
see Fig.~\ref{fig_exp}(b).
These features were reported in Sr$_{2}$IrO$_{4}$ and other Ir oxides. However, their interpretation has been controversial. In early studies~\cite{Kahk14,Yamasaki13} the shoulder feature $A$ was attributed to a charge-transfer final state ($\underline{c}d^6\underline{L}$), while the feature $B$ was interpreted as (unscreened) ionic final states ($\underline{c}d^5$) along a textbook interpretation established for 3$d$ transition-metal oxides (TMOs). In contrast, a recent study~\cite{Horie23} associated the two features with Ir$^{3+}$ and Ir$^{4+}$ valence states. A similar interpretation based on a mixture of different Ir valencies has also been applied to other Ir oxides~\cite{Zhu18,Wu21}.
The Ir $4f$ spectra in Sr$_{2}$IrO$_{4}$ and Sr$_{3}$Ir$_{2}$O$_{7}$ look rather similar and both 
exhibit a change upon cooling below 
$T_N$, as shown 
in Fig.~\ref{fig_exp}(b).
Remarkably, the change in the Ir $4f$ spectra is distinctly different in
the two compounds with even opposite signs around the feature $A$.
Importantly, the position of the distinct peak in $I_\text{300 K}-I_\text{100 K}$ of Sr$_{2}$IrO$_{4}$, indicated by the gray solid bar, is substantially higher ($\sim 0.5$~eV) than that of Sr$_{3}$Ir$_{2}$O$_{7}$, indicated by the pink solid bar. The observed qualitative difference arises from a disparate evolution of spectral intensities with temperature across $T_N$ in the two compounds. Specifically, the intensity of the feature $A$ is suppressed above $T_N$ in Sr$_{2}$IrO$_{4}$, while a tail ($\sim 61.3$~eV) associated with the feature $A$ emerges above $T_N$ in Sr$_{3}$Ir$_{2}$O$_{7}$.

\begin{figure}[t]
\includegraphics[width=0.98\columnwidth]{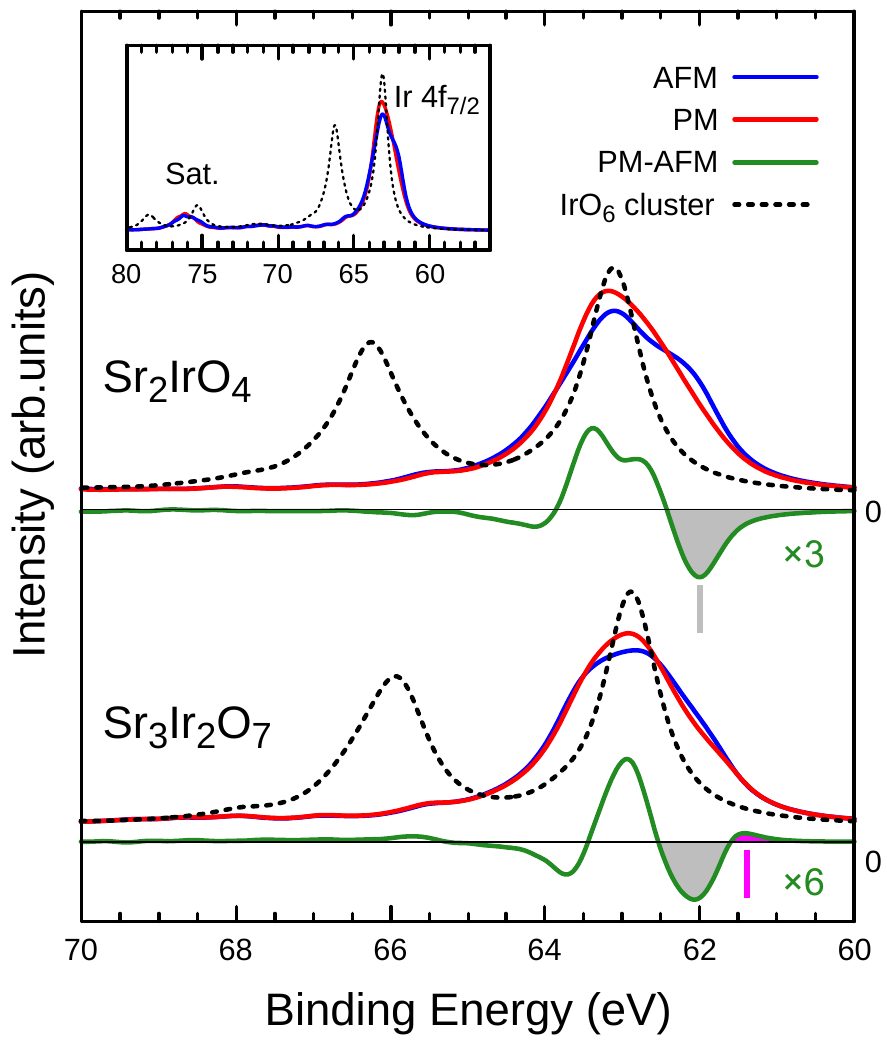}
\caption{Ir 
core-level
spectra calculated by the LDA+DMFT AIM for Sr$_{2}$IrO$_{4}$ (top) and Sr$_{3}$Ir$_{2}$O$_{7}$ (bottom) in the PM (red) and AFM (blue) solutions. The IrO$_6$ cluster-model results for the Ir $4f_{7/2,5/2}$ spectra are shown in the dashed lines. The inset shows the calculated spectra (Sr$_{2}$IrO$_{4}$) in a wide $E_B$ window.}
\label{fig_dmft}
\end{figure}

In the following we use numerical simulations to uncover the microscopic origin
of this behavior. We start with analysis of the temperature and material dependence of Ir 4$f$ spectra of the IrO$_6$ cluster model (dashed lines in Fig.~\ref{fig_dmft}). The model includes the charge transfer from the nearest-neighboring ligands, the local valence-valence and core-valence interactions,  and the spin-orbit coupling in the 4$f$ shell~\footnote{The one-particle Hamiltonian (including hopping integrals and crystal-field energies) in the IrO$_6$ cluster model are extracted from the LDA (tight-binding) Hamiltonian for the experimental crystal structures.}. The cluster model yields sharp structureless 4$f_{7/2}$ and 4$f_{5/2}$ peaks and a weak satellite at high 
$E_B$ ($\sim 75$~eV); see the inset in Fig.~\ref{fig_dmft}. 
The satellite is observed in the experimental spectra
as indicated by  arrows in Fig.~\ref{fig_exp}(a).
Large splitting between the main line and the satellite is due to a strong Ir $5d\ e_g$ -- O 2$p$ hybridization, similar to the origin of the satellite in 2$p$ core-level spectra of early 3$d$ TMOs~\cite{Hariki22,Okada94,Okada93, fn_split}. The  shoulder $A$ is missing in the cluster-model spectrum. 
No core-valence multiplet effects or interference between Ir 4$f_{7/2}$ and 4$f_{5/2}$ excitations are discernible in the cluster-model spectra. This allows us 
to neglect the orbital structure of the core states and compute the 
Ir 4$f_{7/2}$ and 4$f_{5/2}$ spectra separately, which reduces the 
computational effort of LDA+DMFT AIM simulations~\footnote{Note that in this
approximation the Ir 4$f_{7/2}$ and 4$f_{5/2}$ are merely shifted and rescaled
images of one another due to different energy and degeneracy of the
4$f_{7/2}$ and 4$f_{5/2}$ states.}.

Figure~\ref{fig_dmft} also shows the  
Ir core-level spectra
of Sr$_{2}$IrO$_{4}$ and Sr$_{3}$Ir$_{2}$O$_{7}$ obtained with LDA+DMFT AIM. 
The LDA+DMFT AIM treatment produces the  shoulder $A$ in both Sr$_{2}$IrO$_{4}$ and Sr$_{3}$Ir$_{2}$O$_{7}$, in fair agreement with the experimental 4$f_{7/2}$ line. While LDA+DMFT AIM implements the same atomic Hamiltonian of the Ir 5$d$ shell as the cluster model~\cite{Ghiasi19,Hariki17}, it also includes the long-range hopping beyond the nearest-neighboring ligands. The fine structure of the 4$f_{7/2}$ peak can thus be attributed
to a long-range charge-transfer effect, often referred to as {\it nonlocal} screening in the literature~\cite{Veenendaal93,Veenendaal06,Taguchi16book,Hariki17,groot_kotani}.

\begin{figure*}[t]
\includegraphics[width=2.0\columnwidth]{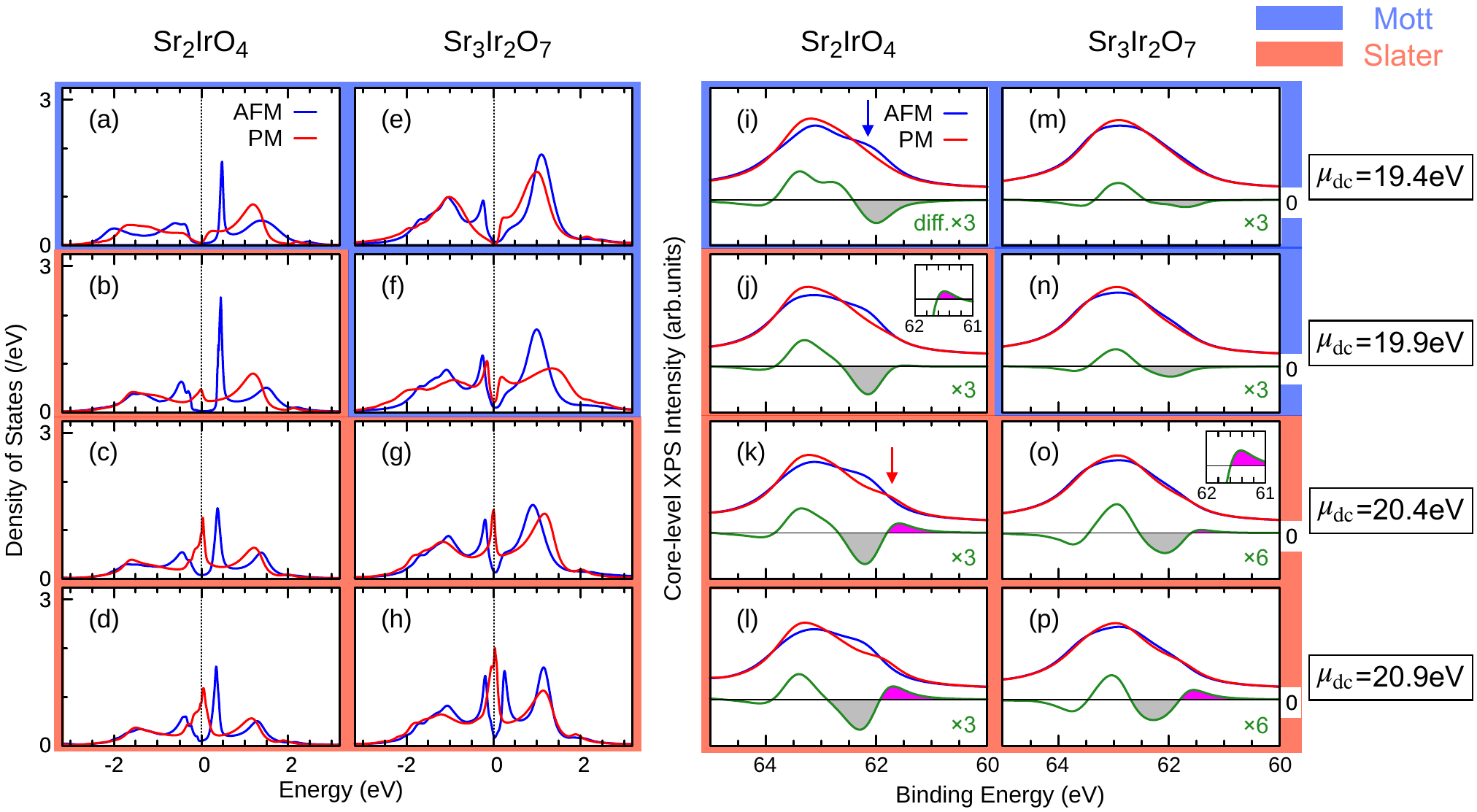}
\caption{LDA+DMFT density of states of the Ir 5$d$ $j_{\rm eff}=1/2$ state and core-level XPS spectra for (a--d, i--l) Sr$_{2}$IrO$_{4}$ and (e--h, m--p) Sr$_{3}$Ir$_{2}$O$_{7}$ calculated for different $\mu_{\rm dc}$ values (from top to bottom panels). The spectra in the PM (red curve) and AFM (blue curve) solutions for each $\mu_{\rm dc}$ value are shown. The results in   blue (red) 
frames 
exhibit the Mott-Hubbard (Slater) type electronic structure with (without) charge gap in the PM solution.
}
\label{fig_dosxps}
\end{figure*}

Next, we address the response of the spectra to the AFM ordering,
characterized by the spectral difference, $I_\text{PM}-I_\text{AFM}$, and its relation to the insulating mechanism. We take advantage of the fact that, in a computer, we can tune between the Mott-Hubbard- and Slater-insulator regimes in both studied materials
by changing the $E_B$ of the Ir $5d$ states away from its realistic value. 
In practice, this is achieved 
by varying 
$\mu_{\rm dc}$.
The LDA+DMFT valence-band spectra for the two regimes are shown 
in Figs.~\ref{fig_dosxps}(a)--\ref{fig_dosxps}(h).
In the Mott-Hubbard regime for Sr$_{2}$IrO$_{4}$ [$\mu_{\rm dc}=19.4$~eV in Fig.\ref{fig_dosxps}(a)], the charge gap at $E_F$ survives in the PM
 insulating (PMI) solution. 
The gap size is slightly increased in the AFM insulating (AFI) solution compared with the PMI one, 
as in the single-band Hubbard model~\cite{Sangiovanni06} and consistent with previous DMFT studies for Sr$_{2}$IrO$_{4}$~\cite{Arita12,Martins2011,Li13,Zhang2013,Lenz2019}. The charge gap collapses in the PM metallic (PMM)
solution in the Slater-insulator regime for Sr$_{3}$Ir$_{2}$O$_{7}$ [$\mu_{\rm dc}=20.4$~eV in Fig.~\ref{fig_dosxps}(g)].  
The results for $I_\text{PM}-I_\text{AFM}$ in 
Figs.~\ref{fig_dosxps}(i)--~\ref{fig_dosxps}(p)
show
that 
the core-level HAXPES spectra reflect
the insulating mechanism rather than a specific material;
i.e.,~similar spectral changes are obtained in both crystal structures of Sr$_{2}$IrO$_{4}$ and Sr$_{3}$Ir$_{2}$O$_{7}$ across the Mott-Hubbard and Slater regimes.
In particular, the spectral weight of the feature $A$ decreases in a Mott-Hubbard case upon the transition from AFI to PMI phase, indicated by the blue arrow in Fig.~\ref{fig_dosxps}(i), resulting in a peak with negative sign in $I_\text{PM}-I_\text{AFM}$. In a Slater  case,
transitioning
from AFI to PMM phase, the feature $A$ increases accompanied by the broad band feature extending to low $E_B$, indicated by the red arrow in Fig.~\ref{fig_dosxps}(k), that produces  a positive sign in $I_\text{PM}-I_\text{AFM}$. 
These behaviors in Fig.~\ref{fig_dosxps}(i) for Sr$_{2}$IrO$_{4}$ and Fig.~\ref{fig_dosxps}(o) for Sr$_{3}$Ir$_{2}$O$_{7}$ well reproduce the observed spectral changes shown in Fig.~\ref{fig_exp}(b).
The spectral evolution with the PMI-to-PMM transition can be found in  Appendix B.
Experimentally, the electrical resistivity of both compounds presents insulating behavior below  $T_N$~\cite{Cao02,Chikara09,L_Li13,Fujiyama12}.
Near $T_N$ the resistivity of Sr$_{3}$Ir$_{2}$O$_{7}$ exhibits a sharp drop,
whereas such a drop is not observed in Sr$_{2}$IrO$_{4}$, suggesting Slater-like and Mott-like character of the gap in the former and latter. 
Our HAXPES result, with the support of the LDA+DMFT simulations, is consistent with the resistivity measurement.

Why does the Ir $4f$ spectrum, in particular the  feature $A$,
respond differently to the AFM transition in the Mott-Hubbard and Slater insulators? 
The lower-energy shoulder feature $A$ is connected with nonlocal screening facilitated
by the states just below $E_F$~\cite{Hariki17,Higashi2021}. Two effects affect
the nonlocal screening in the present context: (i) The nonlocal screening 
is more efficient in a metal than in an insulator due to the presence of 
states close to $E_F$~\footnote{In order to assess whether such states can contribute
to nonlocal screening one should look at hybridization function rather than
an electronic spectral density.}, and
(ii) the nonlocal screening is more efficient in the AFM state than in the PM state since
only the Ir-Ir charge transfer process, in which two electrons with antiparallel spins occupy the $j_{\rm eff}=1/2$ band, is allowed by the Pauli principle and is more preferred in the AFM than in the PM state.
Both (i) and (ii) are active in the Slater regime and the simulations for Sr$_{3}$Ir$_{2}$O$_{7}$ show
that the dominant (i)  
leads to
a larger weight of the feature $A$ in the PM state,  in agreement with the experiment. In contrast,
only
(ii) is active in the Mott-Hubbard regime, resulting in the larger weight of 
the feature $A$ in the AFM state as seen in Sr$_{2}$IrO$_{4}$.
Note that 
the sensitivity of the core-level photoemission spectroscopy to  nonlocal spin-spin correlations or a long-range AFM ordering in a Mott insulator has been proposed theoretically for 3$d$ TMOs~\cite{Hariki13b,Hariki17,Kim04,Haverkort14}, but has been escaping experimental detection so far, except in a recent report for MnO~\cite{Kundu24}. 

Finally we comment on the overestimation of the spectral changes in $I_{\rm PM}-I_{\rm AFM}$ in the LDA+DMFT result in Fig.~\ref{fig_dmft}. This stems primarily from 
a more pronounced suppression of the feature $A$ ($\sim 62$~eV) in the PM solution against the AFM one, compared to the experiment below and above $T_N$, indicating that the deactivation of the nonlocal screening in the PM phase mentioned above is exaggerated in the theory. However, this behavior is not  surprising within the present theory neglecting the short-range AFM correlation. This approximation overcounts the Pauli blocking in the ferromagnetic spin configurations in the time propagation with a core hole and consequently underestimates the intensity of the feature $A$. This observation is consistent with the temperature evolution of the Mn 2$p_{3/2}$ line well above $T_N$ in MnO due to the short-range AFM correlation~\cite{Kundu24}.
Our result thus suggests the potential of core-level HAXPES for studying short-range magnetic correlations in 5$d$ TMOs, and systematic experiments on various compounds, as well as possible extensions of theory to account for them, will be important future challenges.

\section{conclusion}
In summary, we have reported Ir $4f$ core-level HAXPES experiments across the AFM ordering transition in Sr$_{2}$IrO$_{4}$ and Sr$_{3}$Ir$_{2}$O$_{7}$. 
Using LDA+DMFT AIM simulations, we have addressed the origin of the two peak features in the Ir 4$f$ core level, demonstrating the crucial role of nonlocal screening in the analysis of core-level spectra of 5$d$ TMOs, similar to 3$d$ TMOs.
Furthermore, we have explained the microscopic mechanism underlying the observed changes in the HAXPES spectra in response to AFM order, demonstrating that these spectral changes reflect the Slater or Mott-Hubbard character of the AFM insulating state rather than material-specific details.
This broadens the applications of core-level HAXPES as a tool
to study the magnetic and metal-to-insulator transitions in correlated systems.

\begin{acknowledgments} 

We would like to thank S. Yano, T. Hayashida, and S. Miyazaki for supporting the HAXPES experiments.
The HAXPES experiments at SPring-8 were performed  with the approval of 
RIKEN (Proposals No.~20190031 and No.~20200075) 
under the support of JSPS KAKENHI Grants No.~JP19K03753 and No.~JP22K03527.
A.H. was supported by JSPS KAKENHI with Grant No.~JP21K13884, No.~JP21H01003, No.~JP23K03324, No.~JP23H03816, and No.~JP23H03817.
J.~K. was supported by Project No.
CZ.02.01.01/00/22\_008/0004572 of the Programme Johannes Amos Comenius.

\end{acknowledgments}

\appendix
\section{photoemission spectra}

\begin{figure}[bp]
    \includegraphics[width=1\columnwidth]{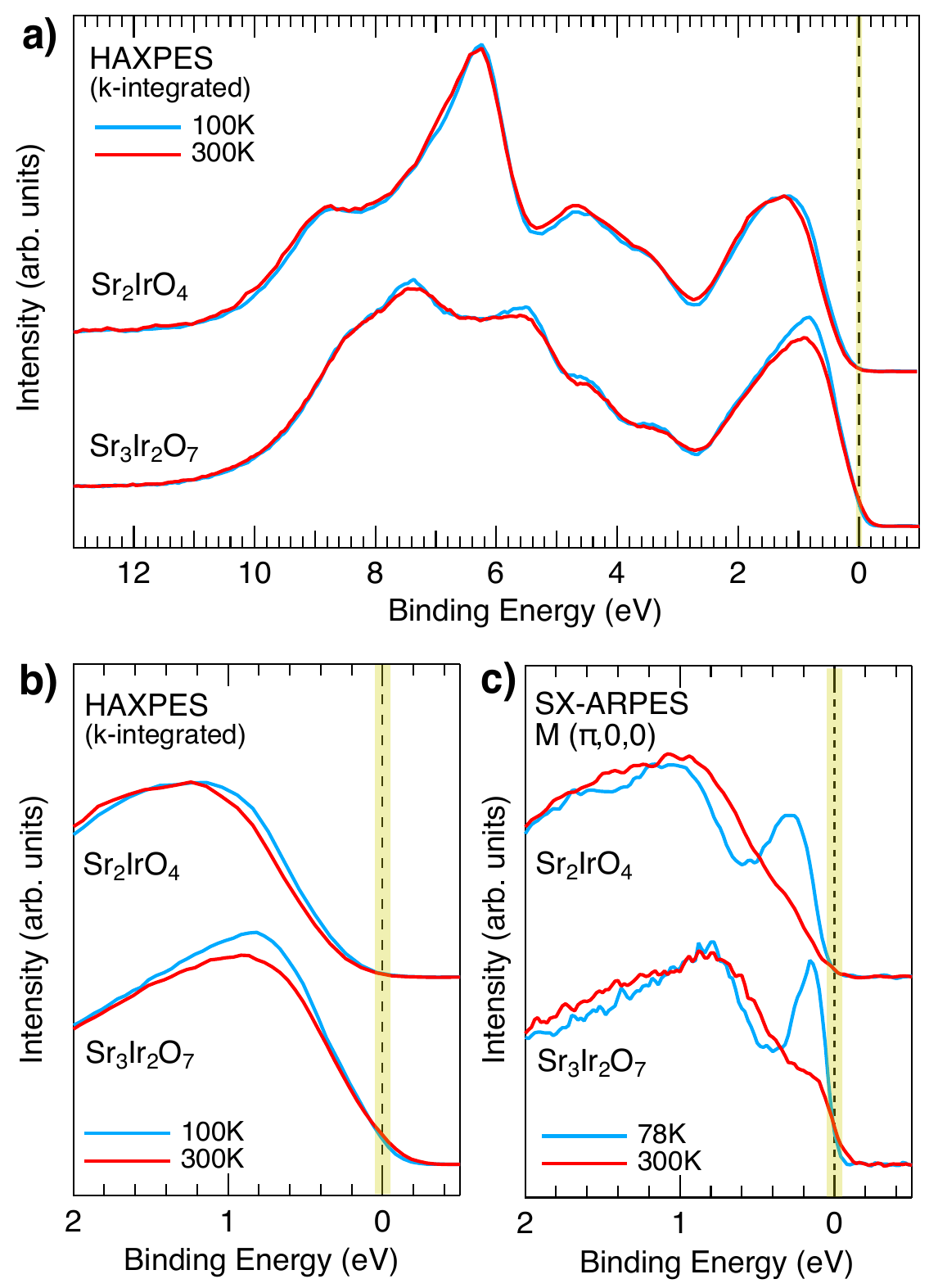}
    \caption{Experimental valence-band spectra of Sr$_2$IrO$_4$ and Sr$_3$Ir$_2$O$_7$ 
    below and above  $T_N$.
    (a) Total and (b) enlarged valence-band HAXPES spectra. 
    SX-ARPES spectra reported in the previous study~\cite{Yamasaki16} are also shown in  (c).}
    \label{fig_VB}
\end{figure}

As reported for many compounds, for example in the strain-induced metal-to-insulator transition system,
the absence or presence of intensity at $E_F$ in the valence-band photoemission spectrum can directly indicate whether a material has a charge gap or not~\cite{Horiba09}.
In this manganite system, it has also been reported that the variation of the intensity at $E_F$ in the valence-band photoemission spectrum corresponds well to that of the intensity of the nonlocal screening structure in the core-level photoemission spectrum.
In contrast, as shown in Figs.~\ref{fig_VB}(a) and ~\ref{fig_VB}(b), for Sr$_{2}$IrO$_{4}$ and Sr$_{3}$Ir$_{2}$O$_{7}$, there is almost no change in intensity not only in the whole valence-band spectrum but also in the vicinity of $E_F$ when comparing between above and below $T_N$. This is due to the small change in electronic structure, 
in addition to the limited energy resolution associated with the use of high-energy photons to increase bulk sensitivity.
Furthermore, the short-range AFM correlation survives even above $T_N$ in the temperature-induced phase transition systems~\cite{Kundu24}, together with the momentum broadening effect on the spectrum with increasing temperature [Fig.~\ref{fig_VB}(c)], making it difficult to discuss the electronic structure change near $E_F$.

\begin{figure}[tbp]
\includegraphics[width=1.0\columnwidth]{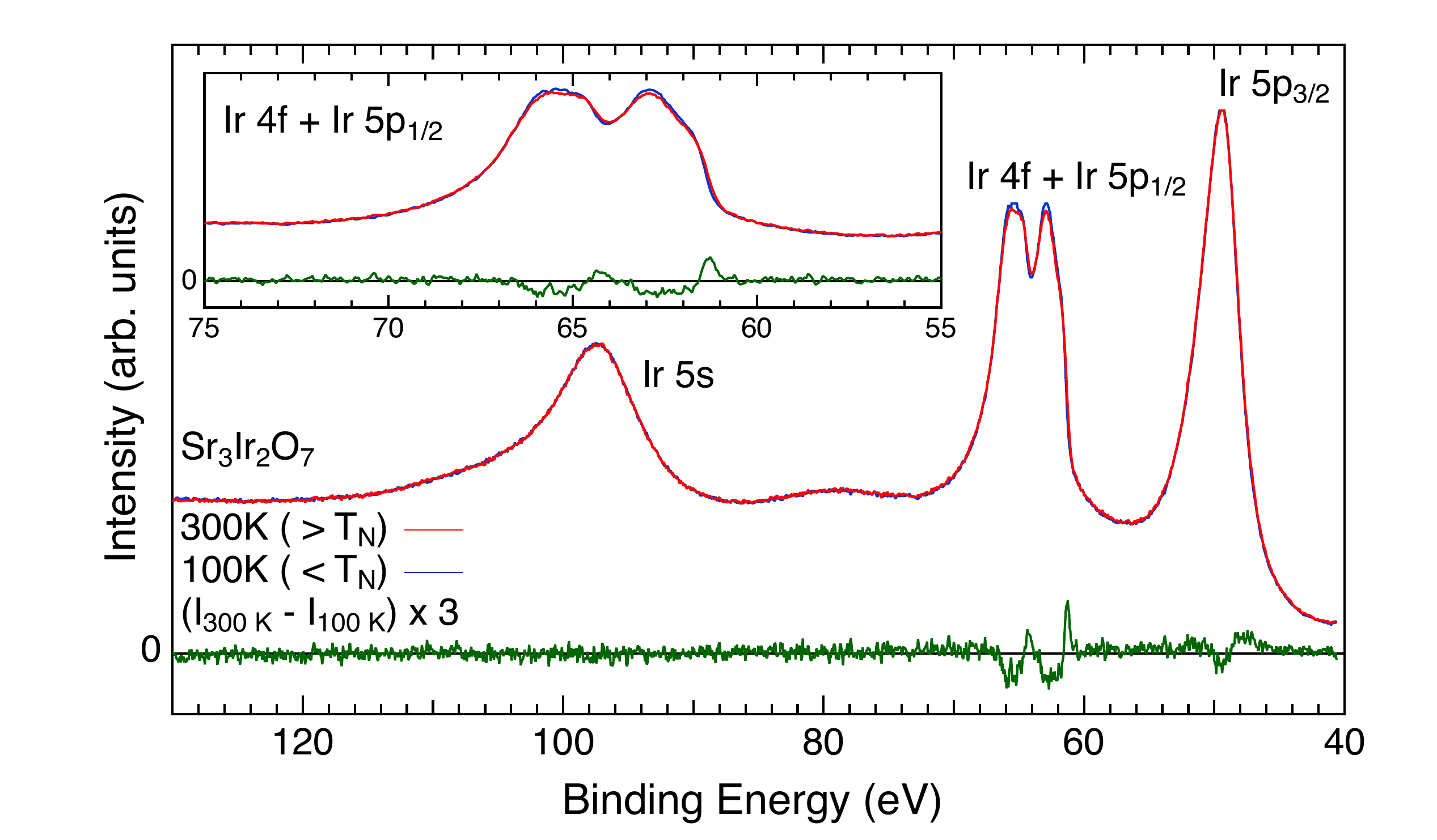}
\caption{HAXPES spectra of Ir core levels at 100 and 300~K in Sr$_{3}$Ir$_{2}$O$_{7}$, shown before background subtraction.
The difference spectrum is also shown.
The inset shows the enlarged spectra around the Ir 4$f$ and 5$p_{1/2}$ core levels.
}
\label{fig_327wide}
\end{figure}

Figure~\ref{fig_327wide} presents the broad-range Ir core-level HAXPES spectra of Sr$_3$Ir$_2$O$_7$ before background subtraction,
together with their temperature variation across $T_N$. Although these supplemental spectra were not recorded with as high a signal-to-noise ratio as the Ir 4$f$ core-level spectra in Fig.~\ref{fig_exp}(b), the difference spectrum between the two temperatures exhibits a distinct feature around the Ir 4$f$ core level, with an intensity well above the noise, similar to the spectrum in Fig.~\ref{fig_exp}(b).
As shown in the inset spectra, this feature does not arise from inaccuracies in the energy calibration of the two spectra but rather from temperature-induced changes in the spectral shape.
The difference spectrum may also include a contribution from  temperature variations in the Ir 5$p_{1/2}$ states, as inferred from the behavior of the 5$p_{3/2}$ core level in Fig.~\ref{fig_327wide}.
However, since the 5$p_{1/2}$ core-level intensity is weaker and its spectral shape  broader than that of the 5$p_{3/2}$ core level, its contribution is expected to be  even smaller~\cite{sm}.
Consequently, the difference spectrum primarily consists  of two contributions:~changes in the shape of the 4$f_{7/2}$ states and corresponding changes in the spin-orbit partner, the 4$f_{5/2}$ states.

\begin{figure}[tbp]
\includegraphics[width=1.0\columnwidth]{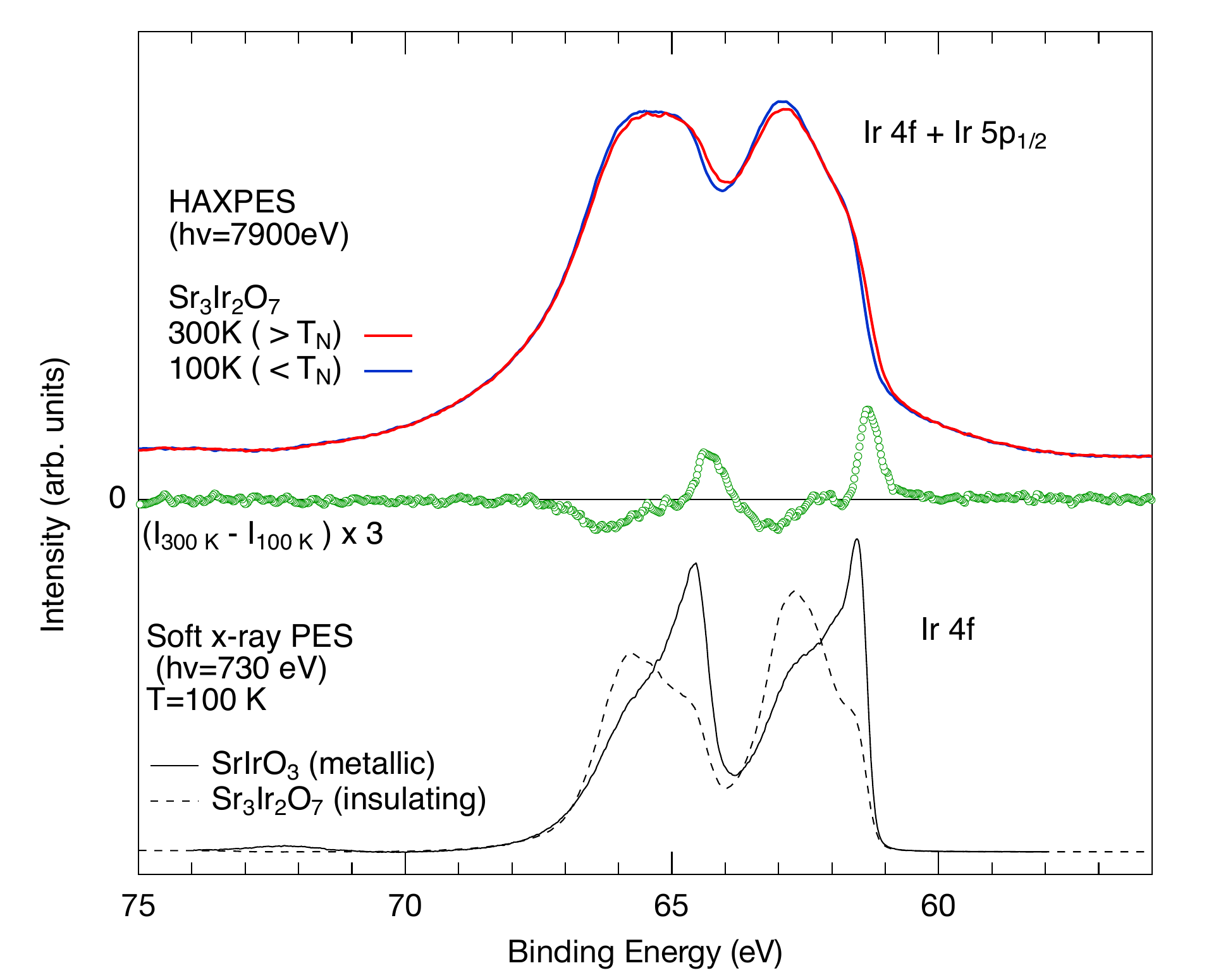}
\caption{HAXPES spectra of Ir 4$f$ core level at 100 and 300~K in Sr$_{3}$Ir$_{2}$O$_{7}$, shown after background subtraction and normalization. 
Soft x-ray photoemission spectra of Sr$_{3}$Ir$_{2}$O$_{7}$ and pseudo-cubic SrIrO$_3$ perovskites are also shown.
}
\label{fig_327and113}
\end{figure}

To further support this spectral interpretation, Fig.~\ref{fig_327and113} presents
the soft x-ray (SX) photoemission spectra of Sr$_{3}$Ir$_{2}$O$_{7}$ and a reference material, pseudocubic SrIrO$_3$~\cite{Yamasaki16}, which behaves as a good metal at 100~K.
In the SX region, the photoionization cross-section ratio $\sigma_{\rm 4\it f}/\sigma_{\rm 5\it p}$ is approximately
fifty times larger than in the hard x-ray region, 
making the SX photoemission spectra  predominantly reflect the Ir 4$f$ states. 
In SrIrO$_3$, a good metal, the low-energy features representing the metallic screening final states are significantly enhanced in both the Ir 4$f_{7/2}$ and 4$f_{5/2}$ lines, reflecting its high metallicity.
In contrast, the insulating Sr$_{3}$Ir$_{2}$O$_{7}$ exhibits broad peaks  at higher binding energies.
The spectral features in these compounds resemble the sharp {\it positive}  and broad {\it negative} 
peaks observed in the difference spectrum in Fig.~\ref{fig_327and113}.
Notably, the two positive-and-negative structures (at 61-64~eV and 64-67~eV) are separated by exactly 
the spin-orbit splitting energy of the Ir 4$f$ core level.

\section{LDA+DMFT and cluster model simulations}

The LDA+DMFT simulation for Ir core-level HAXPES spectra of Sr$_{2}$IrO$_{4}$ and Sr$_{3}$Ir$_{2}$O$_{7}$ proceeds in  two steps. First, a standard LDA+DMFT calculation~\cite{Kotliar06,Georges96} is performed using the same implementation as in Refs.~\onlinecite{Hariki17,Winder20}. Then, the Ir core-level x-ray photoemission spectroscopy (XPS) intensities are computed from the AIM with the optimized LDA+DMFT hybridization densities $\Delta (\omega)$, incorporating the Ir core orbitals and core-valence interaction.

\begin{figure*}[tbp]
\includegraphics[width=1.0\columnwidth]{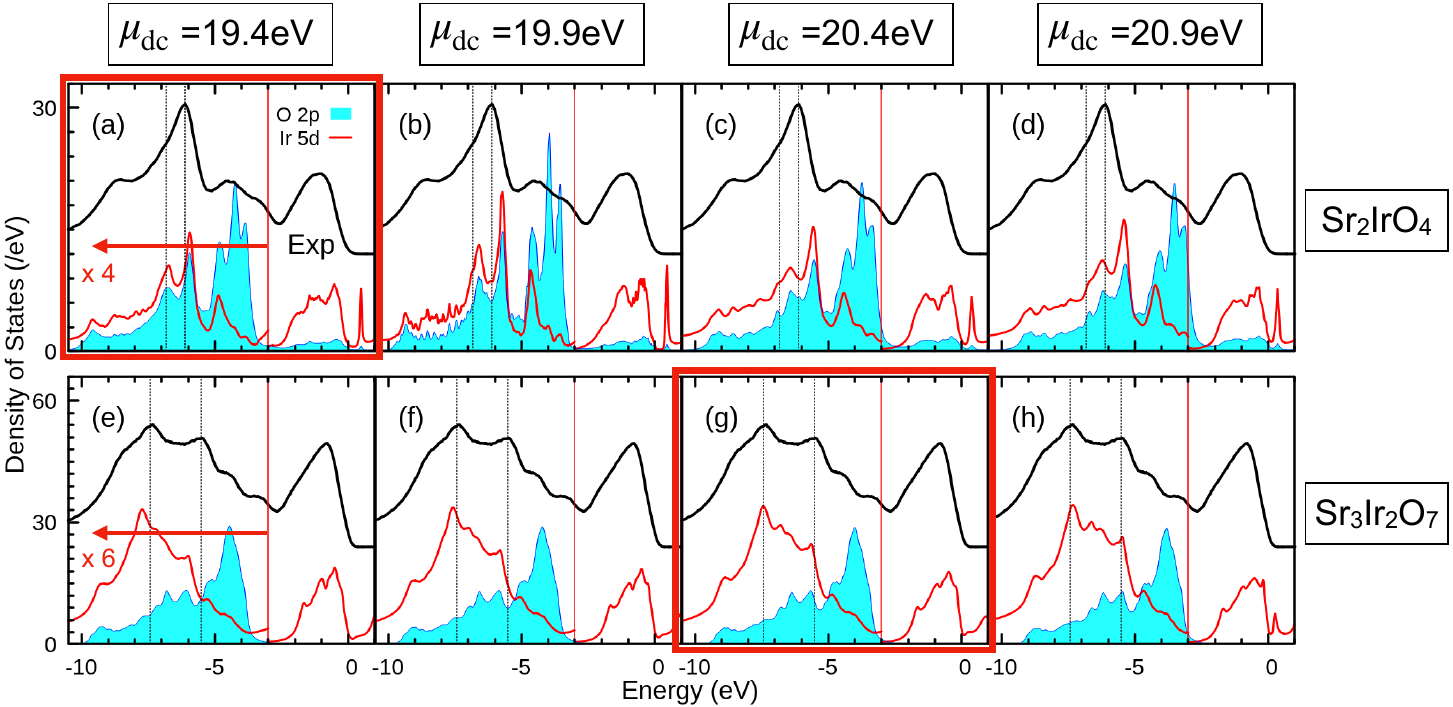}
\caption{LDA+DMFT valence-band spectral intensities of the Ir 5$d$ (red) and O 2$p$ (cyan) states for (a–d, top) Sr$_{2}$IrO$_{4}$ and (e–h, bottom) Sr$_{3}$Ir$_{2}$O$_{7}$ in the AFM phase, calculated for different double-counting values $\mu_{\rm dc}$ (from left to right). The experimental valence-band HAXPES spectra of Sr$_2$IrO$_4$ and Sr$_3$Ir$_2$O$_7$ below the magnetic transition temperature $T_N$ (black curve) are shown for comparison. The intensities of the Ir 5$d$ states below $-$3~eV (indicated by solid red line) are multiplied by a factor of 4 for Sr$_{2}$IrO$_{4}$ and 6 for Sr$_{3}$Ir$_{2}$O$_{7}$ to better visualize the correspondence in the characteristic features with the experimental data (indicated by black dashed lines). The $\mu_{\rm dc}$ values highlighted in red frames represent the best match with the experimental data and are used in the simulations presented in Fig.~\ref{fig_dmft}.
}
\label{fig_dos_af}
\end{figure*}

In the first step, the DFT bands within the LDA for the exchange-correlation potential are obtained with the {\footnotesize{WIEN2K}} package~\cite{wien2k} which employs the augmented plane wave and the local orbital (APW+lo) method.
The experimental crystal structures of Sr$_{2}$IrO$_{4}$ and Sr$_{3}$Ir$_{2}$O$_{7}$ are adopted and spin-orbit coupling is considered in the LDA calculations. 
Subsequently, the DFT bands are projected onto a tight-binding model spanning the Ir 5$d$ and O $2p$ bands using the {\footnotesize{WIEN2WANNIER}} and {\footnotesize{WANNIER90}} codes~\cite{wannier90,wien2wannier}. 
The tight-binding model is augmented by the local electron-electron interaction within the Ir 5$d$ shell. 
The interaction is parametrized by Hubbard $U(=F_0)$ and Hund's $J[=(F_2+F_4)/14]$ parameters, where $F_0$, $F_2$, and $F_4$ are the Slater integrals~\cite{Pavarini1,Pavarini2}.
Consulting with previous DFT-based studies for iridates~\cite{Zhang2013}, we employ $U=4.5$~eV and $J=0.8$~eV. 
The continuous-time quantum Monte Carlo (CT-QMC) method with the hybridization expansion formalism~\cite{Werner2006a,boehnke11,hafermann12} is used to solve the auxiliary AIM in the DMFT self-consistent calculation. 
As described in the main text, the bare Ir $5d$ site energy $\varepsilon_d^{\rm LDA}$ in the LDA tight-binding Hamiltonian is shifted by the double-counting correction $\mu_{\rm dc}$. 
The $\mu_{\rm dc}$ is necessary to subtract the electron-electron interaction effect already present in the LDA results, thereby preventing double counting in the DMFT step. Practically, the $\mu_{\rm dc}$ renormalizes the splitting of the Ir 5$d$ and O 2$p$ states; thus it is related to the charge-transfer energy~\cite{Hariki17,Hariki2020}. Therefore, we could determine its realistic value by comparing the DMFT valence-band spectra with the experimental HAXPES data. Figure~\ref{fig_dos_af} shows the simulated spectra for Sr$_{2}$IrO$_{4}$ and Sr$_{3}$Ir$_{2}$O$_{7}$ in the AFM phase with different $\mu_{\rm dc}$ values. 
In the HAXPES spectra, we observe characteristic features in the high-binding-energy region (indicated by black dashed lines) that represent the bonding features of the O 2$p$ and Ir 5$d$ states, thus serving as a guide to assess the validity of the $\mu_{\rm dc}$ since their binding energies depend on the Ir 5$d$ -- O 2$p$ level splitting, thus $\mu_{\rm dc}$. Indeed, the positions of these features in the simulated spectra are sensitive to variations in $\mu_{\rm dc}$, presenting a systematic shift with $\mu_{\rm dc}$. We found that $\mu_{\rm dc} = 19.4$~eV for Sr$_{2}$IrO$_{4}$ and $20.4$~eV for Sr$_{3}$Ir$_{2}$O$_{7}$ yields good agreement with the experimental data. These values are used in the results (Fig.~2) of the main text.
Once the convergence of the DMFT calculation is achieved, the valence-band spectral intensities and  $\Delta (\omega)$ are computed on the real frequency axis with the self-energy $\Sigma(\omega)$ analytically continued using maximum entropy method~\cite{wang09,jarrell96}.
The  $\Delta (\omega)$ depends on both the orbital and spin indices of the Ir 5$d$ states at the impurity site.

\onecolumngrid\
\twocolumngrid\

In the second step, the AIM with the LDA+DMFT $\Delta (\omega)$ is augmented with the core orbitals and their interaction with the 5$d$ valence electrons at the impurity site. 
The configuration-interaction (CI) solver is adopted to compute core-level XPS intensities using Fermi's golden rule from the LDA+DMFT AIM~\cite{Hariki17,Winder20}.
Similar to high valence or highly covalent 3$d$ TMOs~\cite{Winder20,Hariki22}, due to strong Ir--O covalent bonding, inclusion of a large number of electronic configurations is needed in the CI basis expansion for the studied compounds.
We checked that the Ir 5$d$ occupation obtained in the restricted CI solver is consistent with that in the numerically exact CT-QMC solver (e.g.,~$N^{\rm CI}_d=5.907$ and $N^{\rm QMC}_d=5.884$ for Sr$_{2}$IrO$_{4}$ with $\mu_{\rm dc}=19.4$~eV), and the XPS spectra are well converged for both electronic configurations and the number of discrete bath levels representing the hybridization densities (21~bath levels per spin and orbital are employed in the present CI calculations). 
Because of the extensive basis requirement, full treatment of the 4$f$ core orbital degrees of freedom is not computationally feasible for the studied compounds.
Thus, $s$-type ($l=0$) core orbitals are adopted in the AIM, meaning that the spin-orbit splitting on the Ir 4$f$ core levels and the multipole part in the core-valence (Ir $4f$--$5d$) Coulomb interaction are neglected in the simulated spectra. 
Our determination of the monopole part in the core-valence interaction ($U_{cd}$) will be discussed below.
To complement the spectral interpretation, we also preform spectral calculations for the IrO$_6$ cluster model. 
The cluster model implements the same local Hamiltonian with the LDA+DMFT AIM, while the continuous $\Delta(\omega)$ are replaced by discrete levels composed of the 2$p$ orbitals on the nearest-neighboring O sites within the cluster-model Hamiltonian. The Ir--O hopping parameters are derived from the tight-binding model constructed from the DFT bands~\cite{Ghiasi19}. 
The computationally-cheap cluster model allows a full treatment of the 4$f$ core-orbital degrees of freedom and their multipole interaction with the valence electrons in calculating the spectra.

\begin{figure}[tbp]
\includegraphics[width=1.0\columnwidth]{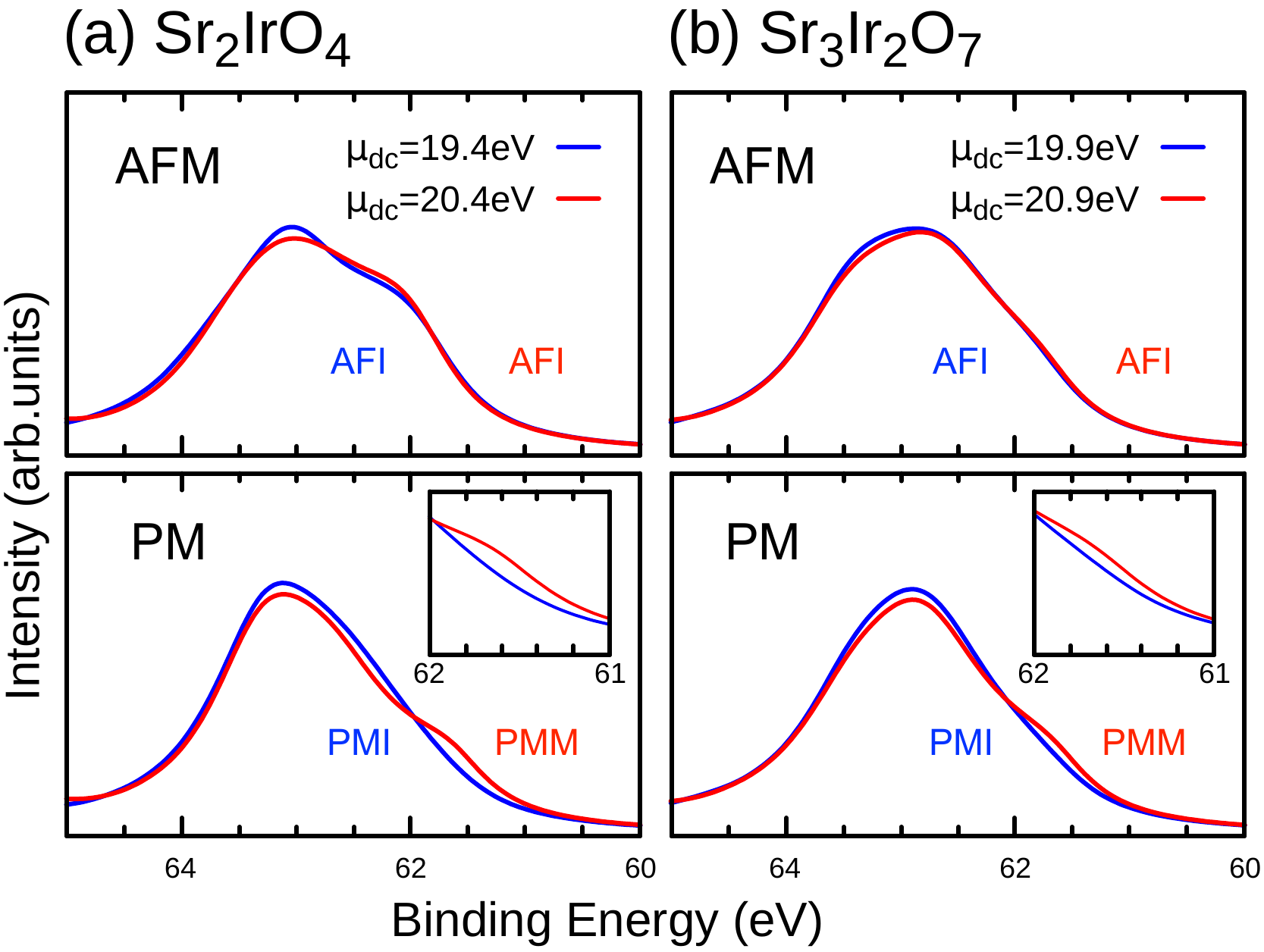}
\caption{Core-level XPS spectra computed by the DMFT method for selected $\mu_{\rm dc}$ values with Mott (blue) and Slater (red) electronic structure
(see also Fig.~\ref{fig_dosxps}) 
for (a) Sr$_{2}$IrO$_{4}$ and (b) Sr$_{3}$Ir$_{2}$O$_{7}$. The inset shows evolution of the metallic-screening feature with the insulator-to-metal transition.}
\label{fig_select_sm}
\end{figure}

In Fig.~\ref{fig_dmft}, we demonstrate that the IrO$_6$ cluster model does not represent the HAXPES spectra, particularly as it fails to capture the fine features in the Ir 4$f_{7/2}$ main line.
An adjustable parameter in the cluster-model Hamiltonian is charge-transfer energy $\Delta_{dp}$, which measures the energy cost to transfer an electron from the O 2$p$ ligand orbitals to the Ir $5d$ orbitals. 
We find
lack of the feature $A$ in the Ir $4f_{7/2}$ main line irrespective of the choice of $\Delta_{dp}$; see Fig.~S2 of the SM~\cite{sm}. 
The IrO$_6$ cluster-model spectra for Sr$_{3}$Ir$_{2}$O$_{7}$ (not shown) are almost identical to those for Sr$_{2}$IrO$_{4}$. 
It should be emphasized that the cluster model explicitly includes the Coulomb multiplet interaction between the Ir 4$f$ core hole and 5$d$ electrons on the x-ray excited site. 
Thus, we can safely exclude the core-valence Coulomb multiplet interaction in the survey of the rich features within the Ir 4$f_{7/2}$ main line.

To supplement Fig.~\ref{fig_dosxps}, Fig.~\ref{fig_select_sm} shows the Ir core-level spectra in the AFM and PM phases of the Mott-Hubbard- and Slater-insulator regimes. The AFM spectra exhibit only minor differences in the two regimes, which is not surprising since the system is insulating in both the regimes under the AFM order. In contrast, the feature $A$ increases accompanied by the broad-band feature in moving from the PMI to PMM phases, i.e.,~from the Mott-Hubbard- to Slater-insulator regimes.

Finally, we describe our choice of core-hole potential parameter $U_{cd}$ in the LDA+DMFT AIM Hamiltonian. For 3$d$ TMOs, an empirical relation for the isotropic part of the core-valence interaction $U_{cd} \approx 1.3 × U_{dd}$ is known~\cite{Zaanen86,Bocquet96,Bocquet92,Matsubara05,Park88}, where $U_{dd}$ is a configuration-averaged valence-valence ($d$--$d$) interaction. 
However, the applicability of this relation to 5$d$ TMOs has not been explored so far. 
To determine the appropriate $U_{cd}$ value, 
we examine
the LDA+DMFT AIM spectra computed with different $U_{cd}$ values. 
We find that the relative intensity of the features $A$ and $B$  in the 4$f_{7/2}$ component is sensitive to the $U_{cd}$ value and the choice of $U_{cd}=1.5\times U_{dd}$ yields a reasonable agreement with the experimental spectrum; see Fig.~S3 of the SM~\cite{sm}.

\bibliography{main}

\begin{thebibliography}{66}%
\makeatletter
\providecommand \@ifxundefined [1]{%
 \@ifx{#1\undefined}
}%
\providecommand \@ifnum [1]{%
 \ifnum #1\expandafter \@firstoftwo
 \else \expandafter \@secondoftwo
 \fi
}%
\providecommand \@ifx [1]{%
 \ifx #1\expandafter \@firstoftwo
 \else \expandafter \@secondoftwo
 \fi
}%
\providecommand \natexlab [1]{#1}%
\providecommand \enquote  [1]{``#1''}%
\providecommand \bibnamefont  [1]{#1}%
\providecommand \bibfnamefont [1]{#1}%
\providecommand \citenamefont [1]{#1}%
\providecommand \href@noop [0]{\@secondoftwo}%
\providecommand \href [0]{\begingroup \@sanitize@url \@href}%
\providecommand \@href[1]{\@@startlink{#1}\@@href}%
\providecommand \@@href[1]{\endgroup#1\@@endlink}%
\providecommand \@sanitize@url [0]{\catcode `\\12\catcode `\$12\catcode
  `\&12\catcode `\#12\catcode `\^12\catcode `\_12\catcode `\%12\relax}%
\providecommand \@@startlink[1]{}%
\providecommand \@@endlink[0]{}%
\providecommand \url  [0]{\begingroup\@sanitize@url \@url }%
\providecommand \@url [1]{\endgroup\@href {#1}{\urlprefix }}%
\providecommand \urlprefix  [0]{URL }%
\providecommand \Eprint [0]{\href }%
\providecommand \doibase [0]{https://doi.org/}%
\providecommand \selectlanguage [0]{\@gobble}%
\providecommand \bibinfo  [0]{\@secondoftwo}%
\providecommand \bibfield  [0]{\@secondoftwo}%
\providecommand \translation [1]{[#1]}%
\providecommand \BibitemOpen [0]{}%
\providecommand \bibitemStop [0]{}%
\providecommand \bibitemNoStop [0]{.\EOS\space}%
\providecommand \EOS [0]{\spacefactor3000\relax}%
\providecommand \BibitemShut  [1]{\csname bibitem#1\endcsname}%
\let\auto@bib@innerbib\@empty
\bibitem [{\citenamefont {King}\ \emph {et~al.}(2013)\citenamefont {King},
  \citenamefont {Takayama}, \citenamefont {Tamai}, \citenamefont {Rozbicki},
  \citenamefont {Walker}, \citenamefont {Shi}, \citenamefont {Patthey},
  \citenamefont {Moore}, \citenamefont {Lu}, \citenamefont {Shen},
  \citenamefont {Takagi},\ and\ \citenamefont {Baumberger}}]{King13}%
  \BibitemOpen
  \bibfield  {author} {\bibinfo {author} {\bibfnamefont {P.~D.~C.}\
  \bibnamefont {King}}, \bibinfo {author} {\bibfnamefont {T.}~\bibnamefont
  {Takayama}}, \bibinfo {author} {\bibfnamefont {A.}~\bibnamefont {Tamai}},
  \bibinfo {author} {\bibfnamefont {E.}~\bibnamefont {Rozbicki}}, \bibinfo
  {author} {\bibfnamefont {S.~M.}\ \bibnamefont {Walker}}, \bibinfo {author}
  {\bibfnamefont {M.}~\bibnamefont {Shi}}, \bibinfo {author} {\bibfnamefont
  {L.}~\bibnamefont {Patthey}}, \bibinfo {author} {\bibfnamefont {R.~G.}\
  \bibnamefont {Moore}}, \bibinfo {author} {\bibfnamefont {D.}~\bibnamefont
  {Lu}}, \bibinfo {author} {\bibfnamefont {K.~M.}\ \bibnamefont {Shen}},
  \bibinfo {author} {\bibfnamefont {H.}~\bibnamefont {Takagi}},\ and\ \bibinfo
  {author} {\bibfnamefont {F.}~\bibnamefont {Baumberger}},\ }\bibfield  {title}
  {\bibinfo {title} {Spectroscopic indications of polaronic behavior of the
  strong spin-orbit insulator {S}r$_{3}${I}r$_{2}${O}$_{7}$},\ }\href
  {https://doi.org/10.1103/PhysRevB.87.241106} {\bibfield  {journal} {\bibinfo
  {journal} {Phys. Rev. B}\ }\textbf {\bibinfo {volume} {87}},\ \bibinfo
  {pages} {241106(R)} (\bibinfo {year} {2013})}\BibitemShut {NoStop}%
\bibitem [{\citenamefont {Yamasaki}\ \emph {et~al.}(2016)\citenamefont
  {Yamasaki}, \citenamefont {Fujiwara}, \citenamefont {Tachibana},
  \citenamefont {Iwasaki}, \citenamefont {Higashino}, \citenamefont {Yoshimi},
  \citenamefont {Nakagawa}, \citenamefont {Nakatani}, \citenamefont {Yamagami},
  \citenamefont {Aratani}, \citenamefont {Kirilmaz}, \citenamefont {Sing},
  \citenamefont {Claessen}, \citenamefont {Watanabe}, \citenamefont
  {Shirakawa}, \citenamefont {Yunoki}, \citenamefont {Naitoh}, \citenamefont
  {Takase}, \citenamefont {Matsuno}, \citenamefont {Takagi}, \citenamefont
  {Sekiyama},\ and\ \citenamefont {Saitoh}}]{Yamasaki16}%
  \BibitemOpen
  \bibfield  {author} {\bibinfo {author} {\bibfnamefont {A.}~\bibnamefont
  {Yamasaki}}, \bibinfo {author} {\bibfnamefont {H.}~\bibnamefont {Fujiwara}},
  \bibinfo {author} {\bibfnamefont {S.}~\bibnamefont {Tachibana}}, \bibinfo
  {author} {\bibfnamefont {D.}~\bibnamefont {Iwasaki}}, \bibinfo {author}
  {\bibfnamefont {Y.}~\bibnamefont {Higashino}}, \bibinfo {author}
  {\bibfnamefont {C.}~\bibnamefont {Yoshimi}}, \bibinfo {author} {\bibfnamefont
  {K.}~\bibnamefont {Nakagawa}}, \bibinfo {author} {\bibfnamefont
  {Y.}~\bibnamefont {Nakatani}}, \bibinfo {author} {\bibfnamefont
  {K.}~\bibnamefont {Yamagami}}, \bibinfo {author} {\bibfnamefont
  {H.}~\bibnamefont {Aratani}}, \bibinfo {author} {\bibfnamefont
  {O.}~\bibnamefont {Kirilmaz}}, \bibinfo {author} {\bibfnamefont
  {M.}~\bibnamefont {Sing}}, \bibinfo {author} {\bibfnamefont {R.}~\bibnamefont
  {Claessen}}, \bibinfo {author} {\bibfnamefont {H.}~\bibnamefont {Watanabe}},
  \bibinfo {author} {\bibfnamefont {T.}~\bibnamefont {Shirakawa}}, \bibinfo
  {author} {\bibfnamefont {S.}~\bibnamefont {Yunoki}}, \bibinfo {author}
  {\bibfnamefont {A.}~\bibnamefont {Naitoh}}, \bibinfo {author} {\bibfnamefont
  {K.}~\bibnamefont {Takase}}, \bibinfo {author} {\bibfnamefont
  {J.}~\bibnamefont {Matsuno}}, \bibinfo {author} {\bibfnamefont
  {H.}~\bibnamefont {Takagi}}, \bibinfo {author} {\bibfnamefont
  {A.}~\bibnamefont {Sekiyama}},\ and\ \bibinfo {author} {\bibfnamefont
  {Y.}~\bibnamefont {Saitoh}},\ }\bibfield  {title} {\bibinfo {title}
  {Three-dimensional electronic structures and the metal-insulator transition
  in {R}uddlesden-{P}opper iridates},\ }\href
  {https://doi.org/10.1103/PhysRevB.94.115103} {\bibfield  {journal} {\bibinfo
  {journal} {Phys. Rev. B}\ }\textbf {\bibinfo {volume} {94}},\ \bibinfo
  {pages} {115103} (\bibinfo {year} {2016})}\BibitemShut {NoStop}%
\bibitem [{\citenamefont {Cao}\ \emph {et~al.}(1998)\citenamefont {Cao},
  \citenamefont {Bolivar}, \citenamefont {McCall}, \citenamefont {Crow},\ and\
  \citenamefont {Guertin}}]{Cao1998}%
  \BibitemOpen
  \bibfield  {author} {\bibinfo {author} {\bibfnamefont {G.}~\bibnamefont
  {Cao}}, \bibinfo {author} {\bibfnamefont {J.}~\bibnamefont {Bolivar}},
  \bibinfo {author} {\bibfnamefont {S.}~\bibnamefont {McCall}}, \bibinfo
  {author} {\bibfnamefont {J.~E.}\ \bibnamefont {Crow}},\ and\ \bibinfo
  {author} {\bibfnamefont {R.~P.}\ \bibnamefont {Guertin}},\ }\bibfield
  {title} {\bibinfo {title} {Weak ferromagnetism, metal-to-nonmetal transition,
  and negative differential resistivity in single-crystal
  $\mathrm{Sr}_{2}\mathrm{IrO}_{4}$},\ }\href
  {https://doi.org/10.1103/PhysRevB.57.R11039} {\bibfield  {journal} {\bibinfo
  {journal} {Phys. Rev. B}\ }\textbf {\bibinfo {volume} {57}},\ \bibinfo
  {pages} {R11039} (\bibinfo {year} {1998})}\BibitemShut {NoStop}%
\bibitem [{\citenamefont {Fujiyama}\ \emph {et~al.}(2012)\citenamefont
  {Fujiyama}, \citenamefont {Ohashi}, \citenamefont {Ohsumi}, \citenamefont
  {Sugimoto}, \citenamefont {Takayama}, \citenamefont {Komesu}, \citenamefont
  {Takata}, \citenamefont {Arima},\ and\ \citenamefont {Takagi}}]{Fujiyama12}%
  \BibitemOpen
  \bibfield  {author} {\bibinfo {author} {\bibfnamefont {S.}~\bibnamefont
  {Fujiyama}}, \bibinfo {author} {\bibfnamefont {K.}~\bibnamefont {Ohashi}},
  \bibinfo {author} {\bibfnamefont {H.}~\bibnamefont {Ohsumi}}, \bibinfo
  {author} {\bibfnamefont {K.}~\bibnamefont {Sugimoto}}, \bibinfo {author}
  {\bibfnamefont {T.}~\bibnamefont {Takayama}}, \bibinfo {author}
  {\bibfnamefont {T.}~\bibnamefont {Komesu}}, \bibinfo {author} {\bibfnamefont
  {M.}~\bibnamefont {Takata}}, \bibinfo {author} {\bibfnamefont
  {T.}~\bibnamefont {Arima}},\ and\ \bibinfo {author} {\bibfnamefont
  {H.}~\bibnamefont {Takagi}},\ }\bibfield  {title} {\bibinfo {title} {Weak
  antiferromagnetism of ${J}_{\rm eff}$= 1/2 band in bilayer iridate
  {S}r$_3${I}r$_2${O}$_7$},\ }\href@noop {} {\bibfield  {journal} {\bibinfo
  {journal} {Phys. Rev. B}\ }\textbf {\bibinfo {volume} {86}},\ \bibinfo
  {pages} {174414} (\bibinfo {year} {2012})}\BibitemShut {NoStop}%
\bibitem [{\citenamefont {Yabashi}\ \emph {et~al.}(2001)\citenamefont
  {Yabashi}, \citenamefont {Tamasaku},\ and\ \citenamefont
  {Ishikawa}}]{Yabashi_01}%
  \BibitemOpen
  \bibfield  {author} {\bibinfo {author} {\bibfnamefont {M.}~\bibnamefont
  {Yabashi}}, \bibinfo {author} {\bibfnamefont {K.}~\bibnamefont {Tamasaku}},\
  and\ \bibinfo {author} {\bibfnamefont {T.}~\bibnamefont {Ishikawa}},\
  }\bibfield  {title} {\bibinfo {title} {Characterization of the transverse
  coherence of hard synchrotron radiation by intensity interferometry},\
  }\href@noop {} {\bibfield  {journal} {\bibinfo  {journal} {Phys. Rev. Lett.}\
  }\textbf {\bibinfo {volume} {87}},\ \bibinfo {pages} {140801} (\bibinfo
  {year} {2001})}\BibitemShut {NoStop}%
\bibitem [{\citenamefont {Fujiwara}\ \emph {et~al.}(2016)\citenamefont
  {Fujiwara}, \citenamefont {Naimen}, \citenamefont {Higashiya}, \citenamefont
  {Kanai}, \citenamefont {Yomosa}, \citenamefont {Yamagami}, \citenamefont
  {Kiss}, \citenamefont {Kadono}, \citenamefont {Imada}, \citenamefont
  {Yamasaki}, \citenamefont {Takase}, \citenamefont {Otsuka}, \citenamefont
  {Shimizu}, \citenamefont {Shingubara}, \citenamefont {Suga}, \citenamefont
  {Yabashi}, \citenamefont {Tamasaku}, \citenamefont {Ishikawa},\ and\
  \citenamefont {Sekiyama}}]{Fujiwara16}%
  \BibitemOpen
  \bibfield  {author} {\bibinfo {author} {\bibfnamefont {H.}~\bibnamefont
  {Fujiwara}}, \bibinfo {author} {\bibfnamefont {S.}~\bibnamefont {Naimen}},
  \bibinfo {author} {\bibfnamefont {A.}~\bibnamefont {Higashiya}}, \bibinfo
  {author} {\bibfnamefont {Y.}~\bibnamefont {Kanai}}, \bibinfo {author}
  {\bibfnamefont {H.}~\bibnamefont {Yomosa}}, \bibinfo {author} {\bibfnamefont
  {K.}~\bibnamefont {Yamagami}}, \bibinfo {author} {\bibfnamefont
  {T.}~\bibnamefont {Kiss}}, \bibinfo {author} {\bibfnamefont {T.}~\bibnamefont
  {Kadono}}, \bibinfo {author} {\bibfnamefont {S.}~\bibnamefont {Imada}},
  \bibinfo {author} {\bibfnamefont {A.}~\bibnamefont {Yamasaki}}, \bibinfo
  {author} {\bibfnamefont {K.}~\bibnamefont {Takase}}, \bibinfo {author}
  {\bibfnamefont {S.}~\bibnamefont {Otsuka}}, \bibinfo {author} {\bibfnamefont
  {T.}~\bibnamefont {Shimizu}}, \bibinfo {author} {\bibfnamefont
  {S.}~\bibnamefont {Shingubara}}, \bibinfo {author} {\bibfnamefont
  {S.}~\bibnamefont {Suga}}, \bibinfo {author} {\bibfnamefont {M.}~\bibnamefont
  {Yabashi}}, \bibinfo {author} {\bibfnamefont {K.}~\bibnamefont {Tamasaku}},
  \bibinfo {author} {\bibfnamefont {T.}~\bibnamefont {Ishikawa}},\ and\
  \bibinfo {author} {\bibfnamefont {A.}~\bibnamefont {Sekiyama}},\ }\bibfield
  {title} {\bibinfo {title} {Polarized hard {X}-ray photoemission system with
  micro-positioning technique for probing ground-state symmetry of strongly
  correlated materials},\ }\href@noop {} {\bibfield  {journal} {\bibinfo
  {journal} {J. Synchrotron Radiat.}\ }\textbf {\bibinfo {volume} {23}},\
  \bibinfo {pages} {735} (\bibinfo {year} {2016})}\BibitemShut {NoStop}%
\bibitem [{\citenamefont {Hariki}\ \emph {et~al.}(2017)\citenamefont {Hariki},
  \citenamefont {Uozumi},\ and\ \citenamefont {Kune\ifmmode~\check{s}\else
  \v{s}\fi{}}}]{Hariki17}%
  \BibitemOpen
  \bibfield  {author} {\bibinfo {author} {\bibfnamefont {A.}~\bibnamefont
  {Hariki}}, \bibinfo {author} {\bibfnamefont {T.}~\bibnamefont {Uozumi}},\
  and\ \bibinfo {author} {\bibfnamefont {J.}~\bibnamefont
  {Kune\ifmmode~\check{s}\else \v{s}\fi{}}},\ }\bibfield  {title} {\bibinfo
  {title} {{LDA}+{DMFT} approach to core-level spectroscopy: Application to
  $3d$ transition metal compounds},\ }\href
  {https://doi.org/10.1103/PhysRevB.96.045111} {\bibfield  {journal} {\bibinfo
  {journal} {Phys. Rev. B}\ }\textbf {\bibinfo {volume} {96}},\ \bibinfo
  {pages} {045111} (\bibinfo {year} {2017})}\BibitemShut {NoStop}%
\bibitem [{\citenamefont {Rahn}\ \emph {et~al.}(2022)\citenamefont {Rahn},
  \citenamefont {Kummer}, \citenamefont {Hariki}, \citenamefont {Ahn},
  \citenamefont {Kune{\v{s}}}, \citenamefont {Amorese}, \citenamefont
  {Denlinger}, \citenamefont {Lu}, \citenamefont {Hashimoto}, \citenamefont
  {Rienks}, \citenamefont {Valvidares}, \citenamefont {Haslbeck}, \citenamefont
  {Byler}, \citenamefont {McClellan}, \citenamefont {Bauer}, \citenamefont
  {Zhu}, \citenamefont {Booth}, \citenamefont {Christianson}, \citenamefont
  {Lawrence}, \citenamefont {Ronning},\ and\ \citenamefont
  {Janoschek}}]{Rahn22}%
  \BibitemOpen
  \bibfield  {author} {\bibinfo {author} {\bibfnamefont {M.~C.}\ \bibnamefont
  {Rahn}}, \bibinfo {author} {\bibfnamefont {K.}~\bibnamefont {Kummer}},
  \bibinfo {author} {\bibfnamefont {A.}~\bibnamefont {Hariki}}, \bibinfo
  {author} {\bibfnamefont {K.-H.}\ \bibnamefont {Ahn}}, \bibinfo {author}
  {\bibfnamefont {J.}~\bibnamefont {Kune{\v{s}}}}, \bibinfo {author}
  {\bibfnamefont {A.}~\bibnamefont {Amorese}}, \bibinfo {author} {\bibfnamefont
  {J.~D.}\ \bibnamefont {Denlinger}}, \bibinfo {author} {\bibfnamefont {D.-H.}\
  \bibnamefont {Lu}}, \bibinfo {author} {\bibfnamefont {M.}~\bibnamefont
  {Hashimoto}}, \bibinfo {author} {\bibfnamefont {E.}~\bibnamefont {Rienks}},
  \bibinfo {author} {\bibfnamefont {M.}~\bibnamefont {Valvidares}}, \bibinfo
  {author} {\bibfnamefont {F.}~\bibnamefont {Haslbeck}}, \bibinfo {author}
  {\bibfnamefont {D.~D.}\ \bibnamefont {Byler}}, \bibinfo {author}
  {\bibfnamefont {K.~J.}\ \bibnamefont {McClellan}}, \bibinfo {author}
  {\bibfnamefont {E.~D.}\ \bibnamefont {Bauer}}, \bibinfo {author}
  {\bibfnamefont {J.~X.}\ \bibnamefont {Zhu}}, \bibinfo {author} {\bibfnamefont
  {C.~H.}\ \bibnamefont {Booth}}, \bibinfo {author} {\bibfnamefont {A.~D.}\
  \bibnamefont {Christianson}}, \bibinfo {author} {\bibfnamefont {J.~M.}\
  \bibnamefont {Lawrence}}, \bibinfo {author} {\bibfnamefont {F.}~\bibnamefont
  {Ronning}},\ and\ \bibinfo {author} {\bibfnamefont {M.}~\bibnamefont
  {Janoschek}},\ }\bibfield  {title} {\bibinfo {title} {Kondo quasiparticle
  dynamics observed by resonant inelastic x-ray scattering},\ }\href
  {https://doi.org/10.1038/s41467-022-33468-6} {\bibfield  {journal} {\bibinfo
  {journal} {Nat. Commun.}\ }\textbf {\bibinfo {volume} {13}},\ \bibinfo
  {pages} {6129} (\bibinfo {year} {2022})}\BibitemShut {NoStop}%
\bibitem [{\citenamefont {Higashi}\ \emph {et~al.}(2021)\citenamefont
  {Higashi}, \citenamefont {Winder}, \citenamefont {Kune\ifmmode~\check{s}\else
  \v{s}\fi{}},\ and\ \citenamefont {Hariki}}]{Higashi2021}%
  \BibitemOpen
  \bibfield  {author} {\bibinfo {author} {\bibfnamefont {K.}~\bibnamefont
  {Higashi}}, \bibinfo {author} {\bibfnamefont {M.}~\bibnamefont {Winder}},
  \bibinfo {author} {\bibfnamefont {J.}~\bibnamefont
  {Kune\ifmmode~\check{s}\else \v{s}\fi{}}},\ and\ \bibinfo {author}
  {\bibfnamefont {A.}~\bibnamefont {Hariki}},\ }\bibfield  {title} {\bibinfo
  {title} {Core-{L}evel {X}-{R}ay {S}pectroscopy of {I}nfinite-{L}ayer
  {N}ickelate: $\mathrm{LDA}+\mathrm{DMFT}$ study},\ }\href
  {https://doi.org/10.1103/PhysRevX.11.041009} {\bibfield  {journal} {\bibinfo
  {journal} {Phys. Rev. X}\ }\textbf {\bibinfo {volume} {11}},\ \bibinfo
  {pages} {041009} (\bibinfo {year} {2021})}\BibitemShut {NoStop}%
\bibitem [{\citenamefont {Zhang}\ \emph {et~al.}(2013)\citenamefont {Zhang},
  \citenamefont {Haule},\ and\ \citenamefont {Vanderbilt}}]{Zhang2013}%
  \BibitemOpen
  \bibfield  {author} {\bibinfo {author} {\bibfnamefont {H.}~\bibnamefont
  {Zhang}}, \bibinfo {author} {\bibfnamefont {K.}~\bibnamefont {Haule}},\ and\
  \bibinfo {author} {\bibfnamefont {D.}~\bibnamefont {Vanderbilt}},\ }\bibfield
   {title} {\bibinfo {title} {Effective ${J}=1/2$ {I}nsulating {S}tate in
  {R}uddlesden-{P}opper {I}ridates: {A}n $\mathrm{LDA}\mathbf{+}\mathrm{DMFT}$
  {S}tudy},\ }\href {https://doi.org/10.1103/PhysRevLett.111.246402} {\bibfield
   {journal} {\bibinfo  {journal} {Phys. Rev. Lett.}\ }\textbf {\bibinfo
  {volume} {111}},\ \bibinfo {pages} {246402} (\bibinfo {year}
  {2013})}\BibitemShut {NoStop}%
\bibitem [{\citenamefont {Kotliar}\ \emph {et~al.}(2006)\citenamefont
  {Kotliar}, \citenamefont {Savrasov}, \citenamefont {Haule}, \citenamefont
  {Oudovenko}, \citenamefont {Parcollet},\ and\ \citenamefont
  {Marianetti}}]{Kotliar06}%
  \BibitemOpen
  \bibfield  {author} {\bibinfo {author} {\bibfnamefont {G.}~\bibnamefont
  {Kotliar}}, \bibinfo {author} {\bibfnamefont {S.~Y.}\ \bibnamefont
  {Savrasov}}, \bibinfo {author} {\bibfnamefont {K.}~\bibnamefont {Haule}},
  \bibinfo {author} {\bibfnamefont {V.~S.}\ \bibnamefont {Oudovenko}}, \bibinfo
  {author} {\bibfnamefont {O.}~\bibnamefont {Parcollet}},\ and\ \bibinfo
  {author} {\bibfnamefont {C.~A.}\ \bibnamefont {Marianetti}},\ }\bibfield
  {title} {\bibinfo {title} {Electronic structure calculations with dynamical
  mean-field theory},\ }\href {https://doi.org/10.1103/RevModPhys.78.865}
  {\bibfield  {journal} {\bibinfo  {journal} {Rev. Mod. Phys.}\ }\textbf
  {\bibinfo {volume} {78}},\ \bibinfo {pages} {865} (\bibinfo {year}
  {2006})}\BibitemShut {NoStop}%
\bibitem [{\citenamefont {Karolak}\ \emph {et~al.}(2010)\citenamefont
  {Karolak}, \citenamefont {Ulm}, \citenamefont {Wehling}, \citenamefont
  {Mazurenko}, \citenamefont {Poteryaev},\ and\ \citenamefont
  {Lichtenstein}}]{Karolak10}%
  \BibitemOpen
  \bibfield  {author} {\bibinfo {author} {\bibfnamefont {M.}~\bibnamefont
  {Karolak}}, \bibinfo {author} {\bibfnamefont {G.}~\bibnamefont {Ulm}},
  \bibinfo {author} {\bibfnamefont {T.}~\bibnamefont {Wehling}}, \bibinfo
  {author} {\bibfnamefont {V.}~\bibnamefont {Mazurenko}}, \bibinfo {author}
  {\bibfnamefont {A.}~\bibnamefont {Poteryaev}},\ and\ \bibinfo {author}
  {\bibfnamefont {A.}~\bibnamefont {Lichtenstein}},\ }\bibfield  {title}
  {\bibinfo {title} {Double counting in {LDA} + {DMFT} - {T}he example of
  {N}i{O}},\ }\href
  {https://doi.org/https://doi.org/10.1016/j.elspec.2010.05.021} {\bibfield
  {journal} {\bibinfo  {journal} {J. Electron Spectrosc. Relat. Phenom.}\
  }\textbf {\bibinfo {volume} {181}},\ \bibinfo {pages} {11 } (\bibinfo {year}
  {2010})}\BibitemShut {NoStop}%
\bibitem [{\citenamefont {Werner}\ \emph {et~al.}(2006)\citenamefont {Werner},
  \citenamefont {Comanac}, \citenamefont {de' Medici}, \citenamefont {Troyer},\
  and\ \citenamefont {Millis}}]{Werner2006a}%
  \BibitemOpen
  \bibfield  {author} {\bibinfo {author} {\bibfnamefont {P.}~\bibnamefont
  {Werner}}, \bibinfo {author} {\bibfnamefont {A.}~\bibnamefont {Comanac}},
  \bibinfo {author} {\bibfnamefont {L.}~\bibnamefont {de' Medici}}, \bibinfo
  {author} {\bibfnamefont {M.}~\bibnamefont {Troyer}},\ and\ \bibinfo {author}
  {\bibfnamefont {A.~J.}\ \bibnamefont {Millis}},\ }\bibfield  {title}
  {\bibinfo {title} {Continuous-{T}ime {S}olver for {Q}uantum {I}mpurity
  {M}odels},\ }\href {https://doi.org/10.1103/PhysRevLett.97.076405} {\bibfield
   {journal} {\bibinfo  {journal} {Phys. Rev. Lett.}\ }\textbf {\bibinfo
  {volume} {97}},\ \bibinfo {pages} {076405} (\bibinfo {year}
  {2006})}\BibitemShut {NoStop}%
\bibitem [{\citenamefont {Kim}\ \emph {et~al.}(2012)\citenamefont {Kim},
  \citenamefont {Choi}, \citenamefont {Kim}, \citenamefont {Mitchell},
  \citenamefont {Jackeli}, \citenamefont {Daghofer}, \citenamefont {van~den
  Brink}, \citenamefont {Khaliullin},\ and\ \citenamefont {Kim}}]{Kim12}%
  \BibitemOpen
  \bibfield  {author} {\bibinfo {author} {\bibfnamefont {J.~W.}\ \bibnamefont
  {Kim}}, \bibinfo {author} {\bibfnamefont {Y.}~\bibnamefont {Choi}}, \bibinfo
  {author} {\bibfnamefont {J.}~\bibnamefont {Kim}}, \bibinfo {author}
  {\bibfnamefont {J.~F.}\ \bibnamefont {Mitchell}}, \bibinfo {author}
  {\bibfnamefont {G.}~\bibnamefont {Jackeli}}, \bibinfo {author} {\bibfnamefont
  {M.}~\bibnamefont {Daghofer}}, \bibinfo {author} {\bibfnamefont
  {J.}~\bibnamefont {van~den Brink}}, \bibinfo {author} {\bibfnamefont
  {G.}~\bibnamefont {Khaliullin}},\ and\ \bibinfo {author} {\bibfnamefont
  {B.~J.}\ \bibnamefont {Kim}},\ }\bibfield  {title} {\bibinfo {title}
  {Dimensionality driven spin-flop transition in layered iridates},\ }\href
  {https://doi.org/10.1103/PhysRevLett.109.037204} {\bibfield  {journal}
  {\bibinfo  {journal} {Phys. Rev. Lett.}\ }\textbf {\bibinfo {volume} {109}},\
  \bibinfo {pages} {037204} (\bibinfo {year} {2012})}\BibitemShut {NoStop}%
\bibitem [{\citenamefont {Kim}\ \emph {et~al.}(2009)\citenamefont {Kim},
  \citenamefont {Ohsumi}, \citenamefont {Komesu}, \citenamefont {Sakai},
  \citenamefont {Morita}, \citenamefont {Takagi},\ and\ \citenamefont
  {Arima}}]{Kim09}%
  \BibitemOpen
  \bibfield  {author} {\bibinfo {author} {\bibfnamefont {B.~J.}\ \bibnamefont
  {Kim}}, \bibinfo {author} {\bibfnamefont {H.}~\bibnamefont {Ohsumi}},
  \bibinfo {author} {\bibfnamefont {T.}~\bibnamefont {Komesu}}, \bibinfo
  {author} {\bibfnamefont {S.}~\bibnamefont {Sakai}}, \bibinfo {author}
  {\bibfnamefont {T.}~\bibnamefont {Morita}}, \bibinfo {author} {\bibfnamefont
  {H.}~\bibnamefont {Takagi}},\ and\ \bibinfo {author} {\bibfnamefont
  {T.}~\bibnamefont {Arima}},\ }\bibfield  {title} {\bibinfo {title}
  {Phase-sensitive observation of a spin-orbital mott state in
  $\mathrm{Sr}_{2}\mathrm{IrO}_{4}$},\ }\href
  {https://doi.org/10.1126/science.1167106} {\bibfield  {journal} {\bibinfo
  {journal} {Science}\ }\textbf {\bibinfo {volume} {323}},\ \bibinfo {pages}
  {1329} (\bibinfo {year} {2009})}\BibitemShut {NoStop}%
\bibitem [{\citenamefont {Jarrell}\ and\ \citenamefont
  {Gubernatis}(1996)}]{jarrell96}%
  \BibitemOpen
  \bibfield  {author} {\bibinfo {author} {\bibfnamefont {M.}~\bibnamefont
  {Jarrell}}\ and\ \bibinfo {author} {\bibfnamefont {J.}~\bibnamefont
  {Gubernatis}},\ }\bibfield  {title} {\bibinfo {title} {Bayesian inference and
  the analytic continuation of imaginary-time quantum {M}onte {C}arlo data},\
  }\href {https://doi.org/http://dx.doi.org/10.1016/0370-1573(95)00074-7}
  {\bibfield  {journal} {\bibinfo  {journal} {Phys. Rep.}\ }\textbf {\bibinfo
  {volume} {269}},\ \bibinfo {pages} {133 } (\bibinfo {year}
  {1996})}\BibitemShut {NoStop}%
\bibitem [{\citenamefont {Ghiasi}\ \emph {et~al.}(2019)\citenamefont {Ghiasi},
  \citenamefont {Hariki}, \citenamefont {Winder}, \citenamefont
  {Kune\ifmmode~\check{s}\else \v{s}\fi{}}, \citenamefont {Regoutz},
  \citenamefont {Lee}, \citenamefont {Hu}, \citenamefont {Rueff},\ and\
  \citenamefont {de~Groot}}]{Ghiasi19}%
  \BibitemOpen
  \bibfield  {author} {\bibinfo {author} {\bibfnamefont {M.}~\bibnamefont
  {Ghiasi}}, \bibinfo {author} {\bibfnamefont {A.}~\bibnamefont {Hariki}},
  \bibinfo {author} {\bibfnamefont {M.}~\bibnamefont {Winder}}, \bibinfo
  {author} {\bibfnamefont {J.}~\bibnamefont {Kune\ifmmode~\check{s}\else
  \v{s}\fi{}}}, \bibinfo {author} {\bibfnamefont {A.}~\bibnamefont {Regoutz}},
  \bibinfo {author} {\bibfnamefont {T.-L.}\ \bibnamefont {Lee}}, \bibinfo
  {author} {\bibfnamefont {Y.}~\bibnamefont {Hu}}, \bibinfo {author}
  {\bibfnamefont {J.-P.}\ \bibnamefont {Rueff}},\ and\ \bibinfo {author}
  {\bibfnamefont {F.~M.~F.}\ \bibnamefont {de~Groot}},\ }\bibfield  {title}
  {\bibinfo {title} {Charge-transfer effect in hard x-ray $1s$ and $2p$
  photoemission spectra: $\mathrm{LDA}+\mathrm{DMFT}$ and cluster-model
  analysis},\ }\href {https://doi.org/10.1103/PhysRevB.100.075146} {\bibfield
  {journal} {\bibinfo  {journal} {Phys. Rev. B}\ }\textbf {\bibinfo {volume}
  {100}},\ \bibinfo {pages} {075146} (\bibinfo {year} {2019})}\BibitemShut
  {NoStop}%
\bibitem [{\citenamefont {Shirley}(1972)}]{Shirley72}%
  \BibitemOpen
  \bibfield  {author} {\bibinfo {author} {\bibfnamefont {D.~A.}\ \bibnamefont
  {Shirley}},\ }\bibfield  {title} {\bibinfo {title} {High-resolution {X}-ray
  photoemission spectrum of the valence bands of gold},\ }\href@noop {}
  {\bibfield  {journal} {\bibinfo  {journal} {Phys. Rev. B}\ }\textbf {\bibinfo
  {volume} {5}},\ \bibinfo {pages} {4709} (\bibinfo {year} {1972})}\BibitemShut
  {NoStop}%
\bibitem [{\citenamefont {Freakley}\ \emph {et~al.}(2017)\citenamefont
  {Freakley}, \citenamefont {Ruiz-Esquius},\ and\ \citenamefont
  {Morgan}}]{Freakley16}%
  \BibitemOpen
  \bibfield  {author} {\bibinfo {author} {\bibfnamefont {S.~J.}\ \bibnamefont
  {Freakley}}, \bibinfo {author} {\bibfnamefont {J.}~\bibnamefont
  {Ruiz-Esquius}},\ and\ \bibinfo {author} {\bibfnamefont {D.~J.}\ \bibnamefont
  {Morgan}},\ }\bibfield  {title} {\bibinfo {title} {The {X}-ray photoelectron
  spectra of {I}r, {I}r{O}$_2$ and {I}r{C}l$_3$ revisited},\ }\href@noop {}
  {\bibfield  {journal} {\bibinfo  {journal} {Surf. Interface Anal.}\ }\textbf
  {\bibinfo {volume} {49}},\ \bibinfo {pages} {794} (\bibinfo {year}
  {2017})}\BibitemShut {NoStop}%
\bibitem [{sm()}]{sm}%
  \BibitemOpen
  \href@noop {} {}\bibinfo {note} {See Supplemental Material for extended data
  on experimental fitting and theoretical parameter studies.}\BibitemShut
  {Stop}%
\bibitem [{Note1()}]{Note1}%
  \BibitemOpen
  \bibinfo {note} {The nonlocal screening effect, discussed in the Ir 4$f$
  spectra below, could also be present in the Ir 5$p_{1/2}$ core-level spectra
  as well. However, the Ir 5$p_{1/2}$ line width is approximately six times
  broader compared to the Ir 4$f$ core level due to large lifetime
  broadening~\cite {sm}. Thus, the temperature dependence should be negligibly
  weak in the Ir 5$p_{1/2}$ spectra.}\BibitemShut {Stop}%
\bibitem [{\citenamefont {Kahk}\ \emph {et~al.}(2014)\citenamefont {Kahk},
  \citenamefont {Poll}, \citenamefont {Oropeza}, \citenamefont {Ablett},
  \citenamefont {C\'eolin}, \citenamefont {Rueff}, \citenamefont {Agrestini},
  \citenamefont {Utsumi}, \citenamefont {Tsuei}, \citenamefont {Liao},
  \citenamefont {Borgatti}, \citenamefont {Panaccione}, \citenamefont
  {Regoutz}, \citenamefont {Egdell}, \citenamefont {Morgan}, \citenamefont
  {Scanlon},\ and\ \citenamefont {Payne}}]{Kahk14}%
  \BibitemOpen
  \bibfield  {author} {\bibinfo {author} {\bibfnamefont {J.~M.}\ \bibnamefont
  {Kahk}}, \bibinfo {author} {\bibfnamefont {C.~G.}\ \bibnamefont {Poll}},
  \bibinfo {author} {\bibfnamefont {F.~E.}\ \bibnamefont {Oropeza}}, \bibinfo
  {author} {\bibfnamefont {J.~M.}\ \bibnamefont {Ablett}}, \bibinfo {author}
  {\bibfnamefont {D.}~\bibnamefont {C\'eolin}}, \bibinfo {author}
  {\bibfnamefont {J.-P.}\ \bibnamefont {Rueff}}, \bibinfo {author}
  {\bibfnamefont {S.}~\bibnamefont {Agrestini}}, \bibinfo {author}
  {\bibfnamefont {Y.}~\bibnamefont {Utsumi}}, \bibinfo {author} {\bibfnamefont
  {K.~D.}\ \bibnamefont {Tsuei}}, \bibinfo {author} {\bibfnamefont {Y.~F.}\
  \bibnamefont {Liao}}, \bibinfo {author} {\bibfnamefont {F.}~\bibnamefont
  {Borgatti}}, \bibinfo {author} {\bibfnamefont {G.}~\bibnamefont
  {Panaccione}}, \bibinfo {author} {\bibfnamefont {A.}~\bibnamefont {Regoutz}},
  \bibinfo {author} {\bibfnamefont {R.~G.}\ \bibnamefont {Egdell}}, \bibinfo
  {author} {\bibfnamefont {B.~J.}\ \bibnamefont {Morgan}}, \bibinfo {author}
  {\bibfnamefont {D.~O.}\ \bibnamefont {Scanlon}},\ and\ \bibinfo {author}
  {\bibfnamefont {D.~J.}\ \bibnamefont {Payne}},\ }\bibfield  {title} {\bibinfo
  {title} {Understanding the {E}lectronic {S}tructure of $\mathrm{IrO}_{2}$
  {U}sing {H}ard-{X}-ray {P}hotoelectron {S}pectroscopy and
  {D}ensity-{F}unctional {T}heory},\ }\href
  {https://doi.org/10.1103/PhysRevLett.112.117601} {\bibfield  {journal}
  {\bibinfo  {journal} {Phys. Rev. Lett.}\ }\textbf {\bibinfo {volume} {112}},\
  \bibinfo {pages} {117601} (\bibinfo {year} {2014})}\BibitemShut {NoStop}%
\bibitem [{\citenamefont {Yamasaki}\ \emph {et~al.}(2014)\citenamefont
  {Yamasaki}, \citenamefont {Kirilmaz}, \citenamefont {Irizawa}, \citenamefont
  {Higashiya}, \citenamefont {Muro}, \citenamefont {Fujiwara}, \citenamefont
  {Pfaff}, \citenamefont {Scheiderer}, \citenamefont {Gabel}, \citenamefont
  {Sing}, \citenamefont {Yabashi}, \citenamefont {Tamasaku}, \citenamefont
  {Hloskovskyy}, \citenamefont {Okabe}, \citenamefont {Yoshida}, \citenamefont
  {Isobe}, \citenamefont {Akimitsu}, \citenamefont {Drube}, \citenamefont
  {Ishikawa}, \citenamefont {Imada}, \citenamefont {Sekiyama}, \citenamefont
  {Claessen},\ and\ \citenamefont {Suga}}]{Yamasaki13}%
  \BibitemOpen
  \bibfield  {author} {\bibinfo {author} {\bibfnamefont {A.}~\bibnamefont
  {Yamasaki}}, \bibinfo {author} {\bibfnamefont {O.}~\bibnamefont {Kirilmaz}},
  \bibinfo {author} {\bibfnamefont {A.}~\bibnamefont {Irizawa}}, \bibinfo
  {author} {\bibfnamefont {A.}~\bibnamefont {Higashiya}}, \bibinfo {author}
  {\bibfnamefont {T.}~\bibnamefont {Muro}}, \bibinfo {author} {\bibfnamefont
  {H.}~\bibnamefont {Fujiwara}}, \bibinfo {author} {\bibfnamefont
  {F.}~\bibnamefont {Pfaff}}, \bibinfo {author} {\bibfnamefont
  {P.}~\bibnamefont {Scheiderer}}, \bibinfo {author} {\bibfnamefont
  {J.}~\bibnamefont {Gabel}}, \bibinfo {author} {\bibfnamefont
  {M.}~\bibnamefont {Sing}}, \bibinfo {author} {\bibfnamefont {M.}~\bibnamefont
  {Yabashi}}, \bibinfo {author} {\bibfnamefont {K.}~\bibnamefont {Tamasaku}},
  \bibinfo {author} {\bibfnamefont {A.}~\bibnamefont {Hloskovskyy}}, \bibinfo
  {author} {\bibfnamefont {H.}~\bibnamefont {Okabe}}, \bibinfo {author}
  {\bibfnamefont {H.}~\bibnamefont {Yoshida}}, \bibinfo {author} {\bibfnamefont
  {M.}~\bibnamefont {Isobe}}, \bibinfo {author} {\bibfnamefont
  {J.}~\bibnamefont {Akimitsu}}, \bibinfo {author} {\bibfnamefont
  {W.}~\bibnamefont {Drube}}, \bibinfo {author} {\bibfnamefont
  {T.}~\bibnamefont {Ishikawa}}, \bibinfo {author} {\bibfnamefont
  {S.}~\bibnamefont {Imada}}, \bibinfo {author} {\bibfnamefont
  {A.}~\bibnamefont {Sekiyama}}, \bibinfo {author} {\bibfnamefont
  {R.}~\bibnamefont {Claessen}},\ and\ \bibinfo {author} {\bibfnamefont
  {S.}~\bibnamefont {Suga}},\ }\bibfield  {title} {\bibinfo {title}
  {Spin-{O}rbit-{C}oupling-{I}nduced $j_{\rm eff}$ {S}tates in {P}erovskite
  {I}ridates {S}tudied by {P}hotoemission {S}pectroscopy},\ }\href
  {https://doi.org/10.7566/JPSCP.3.013001} {\bibfield  {journal} {\bibinfo
  {journal} {JPS Conf. Proc.}\ }\textbf {\bibinfo {volume} {3}},\ \bibinfo
  {pages} {013001} (\bibinfo {year} {2014})}\BibitemShut {NoStop}%
\bibitem [{\citenamefont {Horie}\ \emph {et~al.}(2023)\citenamefont {Horie},
  \citenamefont {Matsushita}, \citenamefont {Kawamura}, \citenamefont {Hase},
  \citenamefont {Horigane}, \citenamefont {Momono}, \citenamefont {Takeuchi},
  \citenamefont {Tanaka}, \citenamefont {Tomita}, \citenamefont {Hashimoto},
  \citenamefont {Kobayashi}, \citenamefont {Haruyama}, \citenamefont {Daimon},
  \citenamefont {Morikawa}, \citenamefont {Taguchi},\ and\ \citenamefont
  {Akimitsu}}]{Horie23}%
  \BibitemOpen
  \bibfield  {author} {\bibinfo {author} {\bibfnamefont {R.}~\bibnamefont
  {Horie}}, \bibinfo {author} {\bibfnamefont {T.}~\bibnamefont {Matsushita}},
  \bibinfo {author} {\bibfnamefont {S.}~\bibnamefont {Kawamura}}, \bibinfo
  {author} {\bibfnamefont {T.}~\bibnamefont {Hase}}, \bibinfo {author}
  {\bibfnamefont {K.}~\bibnamefont {Horigane}}, \bibinfo {author}
  {\bibfnamefont {H.}~\bibnamefont {Momono}}, \bibinfo {author} {\bibfnamefont
  {S.}~\bibnamefont {Takeuchi}}, \bibinfo {author} {\bibfnamefont
  {M.}~\bibnamefont {Tanaka}}, \bibinfo {author} {\bibfnamefont
  {H.}~\bibnamefont {Tomita}}, \bibinfo {author} {\bibfnamefont
  {Y.}~\bibnamefont {Hashimoto}}, \bibinfo {author} {\bibfnamefont
  {K.}~\bibnamefont {Kobayashi}}, \bibinfo {author} {\bibfnamefont
  {Y.}~\bibnamefont {Haruyama}}, \bibinfo {author} {\bibfnamefont
  {H.}~\bibnamefont {Daimon}}, \bibinfo {author} {\bibfnamefont
  {Y.}~\bibnamefont {Morikawa}}, \bibinfo {author} {\bibfnamefont
  {M.}~\bibnamefont {Taguchi}},\ and\ \bibinfo {author} {\bibfnamefont
  {J.}~\bibnamefont {Akimitsu}},\ }\bibfield  {title} {\bibinfo {title} {Origin
  of {U}nexpected {I}r$^{3+}$ in a {S}uperconducting {C}andidate
  {S}r$_2${I}r{O}$_4$ {S}ystem {A}nalyzed by {P}hotoelectron {H}olography},\
  }\href {https://doi.org/10.1021/acs.inorgchem.2c03788} {\bibfield  {journal}
  {\bibinfo  {journal} {Inorg. Chem.}\ }\textbf {\bibinfo {volume} {62}},\
  \bibinfo {pages} {10897} (\bibinfo {year} {2023})}\BibitemShut {NoStop}%
\bibitem [{\citenamefont {Zhu}\ \emph {et~al.}(2018)\citenamefont {Zhu},
  \citenamefont {Liu}, \citenamefont {Cheng}, \citenamefont {Li}, \citenamefont
  {Dong},\ and\ \citenamefont {Wang}}]{Zhu18}%
  \BibitemOpen
  \bibfield  {author} {\bibinfo {author} {\bibfnamefont {C.}~\bibnamefont
  {Zhu}}, \bibinfo {author} {\bibfnamefont {S.}~\bibnamefont {Liu}}, \bibinfo
  {author} {\bibfnamefont {J.}~\bibnamefont {Cheng}}, \bibinfo {author}
  {\bibfnamefont {B.}~\bibnamefont {Li}}, \bibinfo {author} {\bibfnamefont
  {P.}~\bibnamefont {Dong}},\ and\ \bibinfo {author} {\bibfnamefont
  {Z.}~\bibnamefont {Wang}},\ }\bibfield  {title} {\bibinfo {title}
  {Non-monotonic effect of the electronic transport and magnetic properties in
  a $\mathrm{Sm}$-doped $\mathrm{Sr}_{2-x} \mathrm{Sm}_x\mathrm{IrO}_4$
  system},\ }\href {https://doi.org/10.1209/0295-5075/124/17004} {\bibfield
  {journal} {\bibinfo  {journal} {Europhys. Lett.}\ }\textbf {\bibinfo {volume}
  {124}},\ \bibinfo {pages} {17004} (\bibinfo {year} {2018})}\BibitemShut
  {NoStop}%
\bibitem [{\citenamefont {Wu}\ \emph {et~al.}(2021)\citenamefont {Wu},
  \citenamefont {Li}, \citenamefont {Li},\ and\ \citenamefont {Xie}}]{Wu21}%
  \BibitemOpen
  \bibfield  {author} {\bibinfo {author} {\bibfnamefont {Y.}~\bibnamefont
  {Wu}}, \bibinfo {author} {\bibfnamefont {M.}~\bibnamefont {Li}}, \bibinfo
  {author} {\bibfnamefont {X.}~\bibnamefont {Li}},\ and\ \bibinfo {author}
  {\bibfnamefont {J.}~\bibnamefont {Xie}},\ }\bibfield  {title} {\bibinfo
  {title} {Evolution of structural, magnetic, and electrical transport
  properties in $\mathrm{Ru}$-doped pyrochlore iridate
  $\mathrm{Eu}_2\mathrm{Ir}_2\mathrm{O}_7$},\ }\href
  {https://doi.org/10.1007/s10909-020-02524-0} {\bibfield  {journal} {\bibinfo
  {journal} {J. Low Temp. Phys.}\ }\textbf {\bibinfo {volume} {202}},\ \bibinfo
  {pages} {48} (\bibinfo {year} {2021})}\BibitemShut {NoStop}%
\bibitem [{Note2()}]{Note2}%
  \BibitemOpen
  \bibinfo {note} {The one-particle Hamiltonian (including hopping integrals
  and crystal-field energies) in the IrO$_6$ cluster model are extracted from
  the LDA (tight-binding) Hamiltonian for the experimental crystal
  structures.}\BibitemShut {Stop}%
\bibitem [{\citenamefont {Hariki}\ \emph {et~al.}(2022)\citenamefont {Hariki},
  \citenamefont {Higashi}, \citenamefont {Yamaguchi}, \citenamefont {Li},
  \citenamefont {Kalha}, \citenamefont {Mascheck}, \citenamefont {Eriksson},
  \citenamefont {Wiell}, \citenamefont {de~Groot},\ and\ \citenamefont
  {Regoutz}}]{Hariki22}%
  \BibitemOpen
  \bibfield  {author} {\bibinfo {author} {\bibfnamefont {A.}~\bibnamefont
  {Hariki}}, \bibinfo {author} {\bibfnamefont {K.}~\bibnamefont {Higashi}},
  \bibinfo {author} {\bibfnamefont {T.}~\bibnamefont {Yamaguchi}}, \bibinfo
  {author} {\bibfnamefont {J.}~\bibnamefont {Li}}, \bibinfo {author}
  {\bibfnamefont {C.}~\bibnamefont {Kalha}}, \bibinfo {author} {\bibfnamefont
  {M.}~\bibnamefont {Mascheck}}, \bibinfo {author} {\bibfnamefont {S.~K.}\
  \bibnamefont {Eriksson}}, \bibinfo {author} {\bibfnamefont {T.}~\bibnamefont
  {Wiell}}, \bibinfo {author} {\bibfnamefont {F.~M.~F.}\ \bibnamefont
  {de~Groot}},\ and\ \bibinfo {author} {\bibfnamefont {A.}~\bibnamefont
  {Regoutz}},\ }\bibfield  {title} {\bibinfo {title} {Satellites in the {T}i
  $1s$ core level spectra of $\mathrm{SrTiO}_{3}$ and $\mathrm{TiO}_{2}$},\
  }\href {https://doi.org/10.1103/PhysRevB.106.205138} {\bibfield  {journal}
  {\bibinfo  {journal} {Phys. Rev. B}\ }\textbf {\bibinfo {volume} {106}},\
  \bibinfo {pages} {205138} (\bibinfo {year} {2022})}\BibitemShut {NoStop}%
\bibitem [{\citenamefont {Okada}\ \emph {et~al.}(1994)\citenamefont {Okada},
  \citenamefont {Uozumi},\ and\ \citenamefont {Kotani}}]{Okada94}%
  \BibitemOpen
  \bibfield  {author} {\bibinfo {author} {\bibfnamefont {K.}~\bibnamefont
  {Okada}}, \bibinfo {author} {\bibfnamefont {T.}~\bibnamefont {Uozumi}},\ and\
  \bibinfo {author} {\bibfnamefont {A.}~\bibnamefont {Kotani}},\ }\bibfield
  {title} {\bibinfo {title} {Split-off state formation in the final state of
  photoemission in {T}i compounds},\ }\href@noop {} {\bibfield  {journal}
  {\bibinfo  {journal} {J. Phys. Soc. Jpn.}\ }\textbf {\bibinfo {volume}
  {63}},\ \bibinfo {pages} {3176} (\bibinfo {year} {1994})}\BibitemShut
  {NoStop}%
\bibitem [{\citenamefont {Okada}\ and\ \citenamefont {Kotani}(1993)}]{Okada93}%
  \BibitemOpen
  \bibfield  {author} {\bibinfo {author} {\bibfnamefont {K.}~\bibnamefont
  {Okada}}\ and\ \bibinfo {author} {\bibfnamefont {A.}~\bibnamefont {Kotani}},\
  }\bibfield  {title} {\bibinfo {title} {Theory of core level {X}-ray
  photoemission and photoabsorption in {T}i compounds},\ }\href
  {https://doi.org/https://doi.org/10.1016/0368-2048(93)80010-J} {\bibfield
  {journal} {\bibinfo  {journal} {J. Electron Spectrosc. Relat. Phenom.}\
  }\textbf {\bibinfo {volume} {62}},\ \bibinfo {pages} {131} (\bibinfo {year}
  {1993})}\BibitemShut {NoStop}%
\bibitem [{fn_()}]{fn_split}%
  \BibitemOpen
  \href@noop {} {}\bibinfo {note} {For the binding energy of the high-energy
  satellite relative to the main Ir 4$f$ features, the key parameter is an
  effective hybridization given by $V_{\rm
  eff}=\sqrt{V_{e_g}^2\times(4-N_{e_g})+V_{t_{2g}}^2\times(6-N_{t_{2g}})}$,
  where $N_{t_{2g}}$ and $N_{e_{g}}$ denote the occupation numbers of the
  respective orbitals in a formal valence count. Since $N_{t_{2g}}=5$ for
  Ir$^{4+}$, the dominant contribution to $V_{\rm eff}$ arises from
  $V_{e_g}$.}\BibitemShut {Stop}%
\bibitem [{Note3()}]{Note3}%
  \BibitemOpen
  \bibinfo {note} {Note that in this approximation the Ir 4$f_{7/2}$ and
  4$f_{5/2}$ are merely shifted and rescaled images of one another due to
  different energy and degeneracy of the 4$f_{7/2}$ and 4$f_{5/2}$
  states.}\BibitemShut {Stop}%
\bibitem [{\citenamefont {van Veenendaal}\ and\ \citenamefont
  {Sawatzky}(1993)}]{Veenendaal93}%
  \BibitemOpen
  \bibfield  {author} {\bibinfo {author} {\bibfnamefont {M.~A.}\ \bibnamefont
  {van Veenendaal}}\ and\ \bibinfo {author} {\bibfnamefont {G.~A.}\
  \bibnamefont {Sawatzky}},\ }\bibfield  {title} {\bibinfo {title} {Nonlocal
  screening effects in 2$p$ x-ray photoemission spectroscopy core-level line
  shapes of transition metal compounds},\ }\href
  {https://doi.org/10.1103/PhysRevLett.70.2459} {\bibfield  {journal} {\bibinfo
   {journal} {Phys. Rev. Lett.}\ }\textbf {\bibinfo {volume} {70}},\ \bibinfo
  {pages} {2459} (\bibinfo {year} {1993})}\BibitemShut {NoStop}%
\bibitem [{\citenamefont {van Veenendaal}(2006)}]{Veenendaal06}%
  \BibitemOpen
  \bibfield  {author} {\bibinfo {author} {\bibfnamefont {M.}~\bibnamefont {van
  Veenendaal}},\ }\bibfield  {title} {\bibinfo {title} {Competition between
  screening channels in core-level x-ray photoemission as a probe of changes in
  the ground-state properties of transition-metal compounds},\ }\href
  {https://doi.org/10.1103/PhysRevB.74.085118} {\bibfield  {journal} {\bibinfo
  {journal} {Phys. Rev. B}\ }\textbf {\bibinfo {volume} {74}},\ \bibinfo
  {pages} {085118} (\bibinfo {year} {2006})}\BibitemShut {NoStop}%
\bibitem [{\citenamefont {Taguchi}\ and\ \citenamefont
  {Panaccione}(2016)}]{Taguchi16book}%
  \BibitemOpen
  \bibfield  {author} {\bibinfo {author} {\bibfnamefont {M.}~\bibnamefont
  {Taguchi}}\ and\ \bibinfo {author} {\bibfnamefont {G.}~\bibnamefont
  {Panaccione}},\ }\bibinfo {title} {Depth-dependence of electron screening,
  charge carriers and correlation: Theory and experiments},\ in\ \href
  {https://doi.org/10.1007/978-3-319-24043-5_9} {\emph {\bibinfo {booktitle}
  {Hard X-ray Photoelectron Spectroscopy (HAXPES)}}},\ \bibinfo {editor}
  {edited by\ \bibinfo {editor} {\bibfnamefont {J.~C.}\ \bibnamefont
  {Woicik}}}\ (\bibinfo  {publisher} {Springer International Publishing},\
  \bibinfo {address} {Cham},\ \bibinfo {year} {2016})\ pp.\ \bibinfo {pages}
  {197--216}\BibitemShut {NoStop}%
\bibitem [{\citenamefont {de~Groot}\ and\ \citenamefont
  {Kotani}(2014)}]{groot_kotani}%
  \BibitemOpen
  \bibfield  {author} {\bibinfo {author} {\bibfnamefont {F.}~\bibnamefont
  {de~Groot}}\ and\ \bibinfo {author} {\bibfnamefont {A.}~\bibnamefont
  {Kotani}},\ }\href {https://doi.org/10.1201/9781420008425} {\emph {\bibinfo
  {title} {Core Level Spectroscopy of Solids}}}\ (\bibinfo  {publisher} {CRC
  Press, Boca Raton, FL},\ \bibinfo {year} {2014})\BibitemShut {NoStop}%
\bibitem [{\citenamefont {Sangiovanni}\ \emph {et~al.}(2006)\citenamefont
  {Sangiovanni}, \citenamefont {Toschi}, \citenamefont {Koch}, \citenamefont
  {Held}, \citenamefont {Capone}, \citenamefont {Castellani}, \citenamefont
  {Gunnarsson}, \citenamefont {Mo}, \citenamefont {Allen}, \citenamefont {Kim},
  \citenamefont {Sekiyama}, \citenamefont {Yamasaki}, \citenamefont {Suga},\
  and\ \citenamefont {Metcalf}}]{Sangiovanni06}%
  \BibitemOpen
  \bibfield  {author} {\bibinfo {author} {\bibfnamefont {G.}~\bibnamefont
  {Sangiovanni}}, \bibinfo {author} {\bibfnamefont {A.}~\bibnamefont {Toschi}},
  \bibinfo {author} {\bibfnamefont {E.}~\bibnamefont {Koch}}, \bibinfo {author}
  {\bibfnamefont {K.}~\bibnamefont {Held}}, \bibinfo {author} {\bibfnamefont
  {M.}~\bibnamefont {Capone}}, \bibinfo {author} {\bibfnamefont
  {C.}~\bibnamefont {Castellani}}, \bibinfo {author} {\bibfnamefont
  {O.}~\bibnamefont {Gunnarsson}}, \bibinfo {author} {\bibfnamefont {S.-K.}\
  \bibnamefont {Mo}}, \bibinfo {author} {\bibfnamefont {J.~W.}\ \bibnamefont
  {Allen}}, \bibinfo {author} {\bibfnamefont {H.-D.}\ \bibnamefont {Kim}},
  \bibinfo {author} {\bibfnamefont {A.}~\bibnamefont {Sekiyama}}, \bibinfo
  {author} {\bibfnamefont {A.}~\bibnamefont {Yamasaki}}, \bibinfo {author}
  {\bibfnamefont {S.}~\bibnamefont {Suga}},\ and\ \bibinfo {author}
  {\bibfnamefont {P.}~\bibnamefont {Metcalf}},\ }\bibfield  {title} {\bibinfo
  {title} {Static versus dynamical mean-field theory of {M}ott
  antiferromagnets},\ }\href {https://doi.org/10.1103/PhysRevB.73.205121}
  {\bibfield  {journal} {\bibinfo  {journal} {Phys. Rev. B}\ }\textbf {\bibinfo
  {volume} {73}},\ \bibinfo {pages} {205121} (\bibinfo {year}
  {2006})}\BibitemShut {NoStop}%
\bibitem [{\citenamefont {Arita}\ \emph {et~al.}(2012)\citenamefont {Arita},
  \citenamefont {Kune\ifmmode~\check{s}\else \v{s}\fi{}}, \citenamefont
  {Kozhevnikov}, \citenamefont {Eguiluz},\ and\ \citenamefont
  {Imada}}]{Arita12}%
  \BibitemOpen
  \bibfield  {author} {\bibinfo {author} {\bibfnamefont {R.}~\bibnamefont
  {Arita}}, \bibinfo {author} {\bibfnamefont {J.}~\bibnamefont
  {Kune\ifmmode~\check{s}\else \v{s}\fi{}}}, \bibinfo {author} {\bibfnamefont
  {A.~V.}\ \bibnamefont {Kozhevnikov}}, \bibinfo {author} {\bibfnamefont
  {A.~G.}\ \bibnamefont {Eguiluz}},\ and\ \bibinfo {author} {\bibfnamefont
  {M.}~\bibnamefont {Imada}},\ }\bibfield  {title} {\bibinfo {title} {Ab initio
  {S}tudies on the {I}nterplay between {S}pin-{O}rbit {I}nteraction and
  {C}oulomb {C}orrelation in $\mathrm{Sr}_{2}\mathrm{IrO}_{4}$ and
  $\mathrm{Ba}_{2}\mathrm{IrO}_{4}$},\ }\href
  {https://doi.org/10.1103/PhysRevLett.108.086403} {\bibfield  {journal}
  {\bibinfo  {journal} {Phys. Rev. Lett.}\ }\textbf {\bibinfo {volume} {108}},\
  \bibinfo {pages} {086403} (\bibinfo {year} {2012})}\BibitemShut {NoStop}%
\bibitem [{\citenamefont {Martins}\ \emph {et~al.}(2011)\citenamefont
  {Martins}, \citenamefont {Aichhorn}, \citenamefont {Vaugier},\ and\
  \citenamefont {Biermann}}]{Martins2011}%
  \BibitemOpen
  \bibfield  {author} {\bibinfo {author} {\bibfnamefont {C.}~\bibnamefont
  {Martins}}, \bibinfo {author} {\bibfnamefont {M.}~\bibnamefont {Aichhorn}},
  \bibinfo {author} {\bibfnamefont {L.}~\bibnamefont {Vaugier}},\ and\ \bibinfo
  {author} {\bibfnamefont {S.}~\bibnamefont {Biermann}},\ }\bibfield  {title}
  {\bibinfo {title} {Reduced {E}ffective {S}pin-{O}rbital {D}egeneracy and
  {S}pin-{O}rbital {O}rdering in {P}aramagnetic {T}ransition-{M}etal {O}xides:
  $\mathrm{Sr}_{2}\mathrm{IrO}_{4}$ versus $\mathrm{Sr}_{2}\mathrm{RhO}_{4}$},\
  }\href {https://doi.org/10.1103/PhysRevLett.107.266404} {\bibfield  {journal}
  {\bibinfo  {journal} {Phys. Rev. Lett.}\ }\textbf {\bibinfo {volume} {107}},\
  \bibinfo {pages} {266404} (\bibinfo {year} {2011})}\BibitemShut {NoStop}%
\bibitem [{\citenamefont {Li}\ \emph {et~al.}(2013{\natexlab{a}})\citenamefont
  {Li}, \citenamefont {Cao}, \citenamefont {Okamoto}, \citenamefont {Yi},
  \citenamefont {Lin}, \citenamefont {Sales}, \citenamefont {Yan},
  \citenamefont {Arita}, \citenamefont {Kune{\v{s}}}, \citenamefont
  {Kozhevnikov}, \citenamefont {Eguiluz}, \citenamefont {Imada}, \citenamefont
  {Gai}, \citenamefont {Pan},\ and\ \citenamefont {Mandrus}}]{Li13}%
  \BibitemOpen
  \bibfield  {author} {\bibinfo {author} {\bibfnamefont {Q.}~\bibnamefont
  {Li}}, \bibinfo {author} {\bibfnamefont {G.}~\bibnamefont {Cao}}, \bibinfo
  {author} {\bibfnamefont {S.}~\bibnamefont {Okamoto}}, \bibinfo {author}
  {\bibfnamefont {J.}~\bibnamefont {Yi}}, \bibinfo {author} {\bibfnamefont
  {W.}~\bibnamefont {Lin}}, \bibinfo {author} {\bibfnamefont {B.~C.}\
  \bibnamefont {Sales}}, \bibinfo {author} {\bibfnamefont {J.}~\bibnamefont
  {Yan}}, \bibinfo {author} {\bibfnamefont {R.}~\bibnamefont {Arita}}, \bibinfo
  {author} {\bibfnamefont {J.}~\bibnamefont {Kune{\v{s}}}}, \bibinfo {author}
  {\bibfnamefont {A.~V.}\ \bibnamefont {Kozhevnikov}}, \bibinfo {author}
  {\bibfnamefont {A.~G.}\ \bibnamefont {Eguiluz}}, \bibinfo {author}
  {\bibfnamefont {M.}~\bibnamefont {Imada}}, \bibinfo {author} {\bibfnamefont
  {Z.}~\bibnamefont {Gai}}, \bibinfo {author} {\bibfnamefont {M.}~\bibnamefont
  {Pan}},\ and\ \bibinfo {author} {\bibfnamefont {D.~G.}\ \bibnamefont
  {Mandrus}},\ }\bibfield  {title} {\bibinfo {title} {Atomically resolved
  spectroscopic study of {S}r$_2${I}r{O}$_4$: Experiment and theory},\ }\href
  {https://doi.org/10.1038/srep03073} {\bibfield  {journal} {\bibinfo
  {journal} {Sci. Rep.}\ }\textbf {\bibinfo {volume} {3}},\ \bibinfo {pages}
  {3073} (\bibinfo {year} {2013}{\natexlab{a}})}\BibitemShut {NoStop}%
\bibitem [{\citenamefont {Lenz}\ \emph {et~al.}(2019)\citenamefont {Lenz},
  \citenamefont {Martins},\ and\ \citenamefont {Biermann}}]{Lenz2019}%
  \BibitemOpen
  \bibfield  {author} {\bibinfo {author} {\bibfnamefont {B.}~\bibnamefont
  {Lenz}}, \bibinfo {author} {\bibfnamefont {C.}~\bibnamefont {Martins}},\ and\
  \bibinfo {author} {\bibfnamefont {S.}~\bibnamefont {Biermann}},\ }\bibfield
  {title} {\bibinfo {title} {Spectral functions of {S}r$_2${I}r{O}$_4$: theory
  versus experiment},\ }\href {https://doi.org/10.1088/1361-648X/ab146a}
  {\bibfield  {journal} {\bibinfo  {journal} {J. Phys. Condens.}\ }\textbf
  {\bibinfo {volume} {31}},\ \bibinfo {pages} {293001} (\bibinfo {year}
  {2019})}\BibitemShut {NoStop}%
\bibitem [{\citenamefont {Cao}\ \emph {et~al.}(2002)\citenamefont {Cao},
  \citenamefont {Xin}, \citenamefont {Alexander}, \citenamefont {Crow},
  \citenamefont {Schlottmann}, \citenamefont {Crawford}, \citenamefont
  {Harlow},\ and\ \citenamefont {Marshall}}]{Cao02}%
  \BibitemOpen
  \bibfield  {author} {\bibinfo {author} {\bibfnamefont {G.}~\bibnamefont
  {Cao}}, \bibinfo {author} {\bibfnamefont {Y.}~\bibnamefont {Xin}}, \bibinfo
  {author} {\bibfnamefont {C.~S.}\ \bibnamefont {Alexander}}, \bibinfo {author}
  {\bibfnamefont {J.~E.}\ \bibnamefont {Crow}}, \bibinfo {author}
  {\bibfnamefont {P.}~\bibnamefont {Schlottmann}}, \bibinfo {author}
  {\bibfnamefont {M.~K.}\ \bibnamefont {Crawford}}, \bibinfo {author}
  {\bibfnamefont {R.~L.}\ \bibnamefont {Harlow}},\ and\ \bibinfo {author}
  {\bibfnamefont {W.}~\bibnamefont {Marshall}},\ }\bibfield  {title} {\bibinfo
  {title} {Anomalous magnetic and transport behavior in the magnetic insulator
  {S}r$_3${I}r$_2${O}$_7$},\ }\href@noop {} {\bibfield  {journal} {\bibinfo
  {journal} {Phys. Rev. B}\ }\textbf {\bibinfo {volume} {66}},\ \bibinfo
  {pages} {214412} (\bibinfo {year} {2002})}\BibitemShut {NoStop}%
\bibitem [{\citenamefont {Chikara}\ \emph {et~al.}(2009)\citenamefont
  {Chikara}, \citenamefont {Korneta}, \citenamefont {Crummett}, \citenamefont
  {DeLong}, \citenamefont {Schlottmann},\ and\ \citenamefont
  {Cao}}]{Chikara09}%
  \BibitemOpen
  \bibfield  {author} {\bibinfo {author} {\bibfnamefont {S.}~\bibnamefont
  {Chikara}}, \bibinfo {author} {\bibfnamefont {O.}~\bibnamefont {Korneta}},
  \bibinfo {author} {\bibfnamefont {W.~P.}\ \bibnamefont {Crummett}}, \bibinfo
  {author} {\bibfnamefont {L.~E.}\ \bibnamefont {DeLong}}, \bibinfo {author}
  {\bibfnamefont {P.}~\bibnamefont {Schlottmann}},\ and\ \bibinfo {author}
  {\bibfnamefont {G.}~\bibnamefont {Cao}},\ }\bibfield  {title} {\bibinfo
  {title} {Giant magnetoelectric effect in the ${J}_\mathrm{eff}=\frac{1}{2}$
  {M}ott insulator {S}r$_{2}${I}r{O}$_{4}$},\ }\href
  {https://doi.org/10.1103/PhysRevB.80.140407} {\bibfield  {journal} {\bibinfo
  {journal} {Phys. Rev. B}\ }\textbf {\bibinfo {volume} {80}},\ \bibinfo
  {pages} {140407(R)} (\bibinfo {year} {2009})}\BibitemShut {NoStop}%
\bibitem [{\citenamefont {Li}\ \emph {et~al.}(2013{\natexlab{b}})\citenamefont
  {Li}, \citenamefont {Kong}, \citenamefont {Qi}, \citenamefont {Jin},
  \citenamefont {Yuan}, \citenamefont {DeLong}, \citenamefont {Schlottmann},\
  and\ \citenamefont {Cao}}]{L_Li13}%
  \BibitemOpen
  \bibfield  {author} {\bibinfo {author} {\bibfnamefont {L.}~\bibnamefont
  {Li}}, \bibinfo {author} {\bibfnamefont {P.~P.}\ \bibnamefont {Kong}},
  \bibinfo {author} {\bibfnamefont {T.~F.}\ \bibnamefont {Qi}}, \bibinfo
  {author} {\bibfnamefont {C.~Q.}\ \bibnamefont {Jin}}, \bibinfo {author}
  {\bibfnamefont {S.~J.}\ \bibnamefont {Yuan}}, \bibinfo {author}
  {\bibfnamefont {L.~E.}\ \bibnamefont {DeLong}}, \bibinfo {author}
  {\bibfnamefont {P.}~\bibnamefont {Schlottmann}},\ and\ \bibinfo {author}
  {\bibfnamefont {G.}~\bibnamefont {Cao}},\ }\bibfield  {title} {\bibinfo
  {title} {Tuning the ${J}_{\mathrm{eff}}=\frac{1}{2}$ insulating state via
  electron doping and pressure in the double-layered iridate
  {S}r$_{3}${I}r$_{2}${O}$_{7}$},\ }\href
  {https://doi.org/10.1103/PhysRevB.87.235127} {\bibfield  {journal} {\bibinfo
  {journal} {Phys. Rev. B}\ }\textbf {\bibinfo {volume} {87}},\ \bibinfo
  {pages} {235127} (\bibinfo {year} {2013}{\natexlab{b}})}\BibitemShut
  {NoStop}%
\bibitem [{Note4()}]{Note4}%
  \BibitemOpen
  \bibinfo {note} {In order to assess whether such states can contribute to
  nonlocal screening one should look at hybridization function rather than an
  electronic spectral density.}\BibitemShut {Stop}%
\bibitem [{\citenamefont {Hariki}\ \emph {et~al.}(2013)\citenamefont {Hariki},
  \citenamefont {Ichinozuka},\ and\ \citenamefont {Uozumi}}]{Hariki13b}%
  \BibitemOpen
  \bibfield  {author} {\bibinfo {author} {\bibfnamefont {A.}~\bibnamefont
  {Hariki}}, \bibinfo {author} {\bibfnamefont {Y.}~\bibnamefont {Ichinozuka}},\
  and\ \bibinfo {author} {\bibfnamefont {T.}~\bibnamefont {Uozumi}},\
  }\bibfield  {title} {\bibinfo {title} {Dynamical {M}ean-{F}ield {A}pproach to
  {N}i 2$p$ {X}-ray {P}hotoemission {S}pectra of {N}i{O}: {A} {R}ole of
  {A}ntiferromagnetic {O}rdering},\ }\href
  {https://doi.org/10.7566/JPSJ.82.043710} {\bibfield  {journal} {\bibinfo
  {journal} {J. Phys. Soc. Jpn.}\ }\textbf {\bibinfo {volume} {82}},\ \bibinfo
  {pages} {043710} (\bibinfo {year} {2013})}\BibitemShut {NoStop}%
\bibitem [{\citenamefont {Kim}\ \emph {et~al.}(2004)\citenamefont {Kim},
  \citenamefont {Noh}, \citenamefont {Kim},\ and\ \citenamefont {Oh}}]{Kim04}%
  \BibitemOpen
  \bibfield  {author} {\bibinfo {author} {\bibfnamefont {H.-D.}\ \bibnamefont
  {Kim}}, \bibinfo {author} {\bibfnamefont {H.-J.}\ \bibnamefont {Noh}},
  \bibinfo {author} {\bibfnamefont {K.~H.}\ \bibnamefont {Kim}},\ and\ \bibinfo
  {author} {\bibfnamefont {S.-J.}\ \bibnamefont {Oh}},\ }\bibfield  {title}
  {\bibinfo {title} {Core-{L}evel {X}-{R}ay {P}hotoemission {S}atellites in
  {R}uthenates: {A} {N}ew {M}echanism {R}evealing {T}he {M}ott {T}ransition},\
  }\href {https://doi.org/10.1103/PhysRevLett.93.126404} {\bibfield  {journal}
  {\bibinfo  {journal} {Phys. Rev. Lett.}\ }\textbf {\bibinfo {volume} {93}},\
  \bibinfo {pages} {126404} (\bibinfo {year} {2004})}\BibitemShut {NoStop}%
\bibitem [{\citenamefont {Haverkort}\ \emph {et~al.}(2014)\citenamefont
  {Haverkort}, \citenamefont {Sangiovanni}, \citenamefont {Hansmann},
  \citenamefont {Toschi}, \citenamefont {Lu},\ and\ \citenamefont
  {Macke}}]{Haverkort14}%
  \BibitemOpen
  \bibfield  {author} {\bibinfo {author} {\bibfnamefont {M.~W.}\ \bibnamefont
  {Haverkort}}, \bibinfo {author} {\bibfnamefont {G.}~\bibnamefont
  {Sangiovanni}}, \bibinfo {author} {\bibfnamefont {P.}~\bibnamefont
  {Hansmann}}, \bibinfo {author} {\bibfnamefont {A.}~\bibnamefont {Toschi}},
  \bibinfo {author} {\bibfnamefont {Y.}~\bibnamefont {Lu}},\ and\ \bibinfo
  {author} {\bibfnamefont {S.}~\bibnamefont {Macke}},\ }\bibfield  {title}
  {\bibinfo {title} {Bands, resonances, edge singularities and excitons in core
  level spectroscopy investigated within the dynamical mean-field theory},\
  }\href {https://doi.org/10.1209/0295-5075/108/57004} {\bibfield  {journal}
  {\bibinfo  {journal} {Europhys. Lett.}\ }\textbf {\bibinfo {volume} {108}},\
  \bibinfo {pages} {57004} (\bibinfo {year} {2014})}\BibitemShut {NoStop}%
\bibitem [{\citenamefont {Kundu}\ \emph {et~al.}(2024)\citenamefont {Kundu},
  \citenamefont {Sheverdyaeva}, \citenamefont {Moras}, \citenamefont {Menon},
  \citenamefont {Mandal},\ and\ \citenamefont {Carbone}}]{Kundu24}%
  \BibitemOpen
  \bibfield  {author} {\bibinfo {author} {\bibfnamefont {A.~K.}\ \bibnamefont
  {Kundu}}, \bibinfo {author} {\bibfnamefont {P.~M.}\ \bibnamefont
  {Sheverdyaeva}}, \bibinfo {author} {\bibfnamefont {P.}~\bibnamefont {Moras}},
  \bibinfo {author} {\bibfnamefont {K.~S.~R.}\ \bibnamefont {Menon}}, \bibinfo
  {author} {\bibfnamefont {S.}~\bibnamefont {Mandal}},\ and\ \bibinfo {author}
  {\bibfnamefont {C.}~\bibnamefont {Carbone}},\ }\bibfield  {title} {\bibinfo
  {title} {Spin-selective evolution of the zhang-rice state in binary
  transition metal oxide $\mathrm{MnO}$ (001) film},\ }\href
  {https://doi.org/10.1103/PhysRevB.109.195111} {\bibfield  {journal} {\bibinfo
   {journal} {Phys. Rev. B}\ }\textbf {\bibinfo {volume} {109}},\ \bibinfo
  {pages} {195111} (\bibinfo {year} {2024})}\BibitemShut {NoStop}%
\bibitem [{\citenamefont {Horiba}\ \emph {et~al.}(2009)\citenamefont {Horiba},
  \citenamefont {Maniwa}, \citenamefont {Chikamatsu}, \citenamefont
  {Yoshimatsu}, \citenamefont {Kumigashira}, \citenamefont {Wadati},
  \citenamefont {Fujimori}, \citenamefont {Ueda}, \citenamefont {Yoshikawa},
  \citenamefont {Ikenaga}, \citenamefont {Kim}, \citenamefont {Kobayashi},\
  and\ \citenamefont {Oshima}}]{Horiba09}%
  \BibitemOpen
  \bibfield  {author} {\bibinfo {author} {\bibfnamefont {K.}~\bibnamefont
  {Horiba}}, \bibinfo {author} {\bibfnamefont {A.}~\bibnamefont {Maniwa}},
  \bibinfo {author} {\bibfnamefont {A.}~\bibnamefont {Chikamatsu}}, \bibinfo
  {author} {\bibfnamefont {K.}~\bibnamefont {Yoshimatsu}}, \bibinfo {author}
  {\bibfnamefont {H.}~\bibnamefont {Kumigashira}}, \bibinfo {author}
  {\bibfnamefont {H.}~\bibnamefont {Wadati}}, \bibinfo {author} {\bibfnamefont
  {A.}~\bibnamefont {Fujimori}}, \bibinfo {author} {\bibfnamefont
  {S.}~\bibnamefont {Ueda}}, \bibinfo {author} {\bibfnamefont {H.}~\bibnamefont
  {Yoshikawa}}, \bibinfo {author} {\bibfnamefont {E.}~\bibnamefont {Ikenaga}},
  \bibinfo {author} {\bibfnamefont {J.~J.}\ \bibnamefont {Kim}}, \bibinfo
  {author} {\bibfnamefont {K.}~\bibnamefont {Kobayashi}},\ and\ \bibinfo
  {author} {\bibfnamefont {M.}~\bibnamefont {Oshima}},\ }\bibfield  {title}
  {\bibinfo {title} {Pressure-induced change in the electronic structure of
  epitaxially strained $\mathrm{La}_{1-x}\mathrm{Sr}_{x}\mathrm{MnO}_{3}$ thin
  films},\ }\href {https://doi.org/10.1103/PhysRevB.80.132406} {\bibfield
  {journal} {\bibinfo  {journal} {Phys. Rev. B}\ }\textbf {\bibinfo {volume}
  {80}},\ \bibinfo {pages} {132406} (\bibinfo {year} {2009})}\BibitemShut
  {NoStop}%
\bibitem [{\citenamefont {Georges}\ \emph {et~al.}(1996)\citenamefont
  {Georges}, \citenamefont {Kotliar}, \citenamefont {Krauth},\ and\
  \citenamefont {Rozenberg}}]{Georges96}%
  \BibitemOpen
  \bibfield  {author} {\bibinfo {author} {\bibfnamefont {A.}~\bibnamefont
  {Georges}}, \bibinfo {author} {\bibfnamefont {G.}~\bibnamefont {Kotliar}},
  \bibinfo {author} {\bibfnamefont {W.}~\bibnamefont {Krauth}},\ and\ \bibinfo
  {author} {\bibfnamefont {M.~J.}\ \bibnamefont {Rozenberg}},\ }\bibfield
  {title} {\bibinfo {title} {Dynamical mean-field theory of strongly correlated
  fermion systems and the limit of infinite dimensions},\ }\href
  {https://doi.org/10.1103/RevModPhys.68.13} {\bibfield  {journal} {\bibinfo
  {journal} {Rev. Mod. Phys.}\ }\textbf {\bibinfo {volume} {68}},\ \bibinfo
  {pages} {13} (\bibinfo {year} {1996})}\BibitemShut {NoStop}%
\bibitem [{\citenamefont {Winder}\ \emph {et~al.}(2020)\citenamefont {Winder},
  \citenamefont {Hariki},\ and\ \citenamefont {Kune\ifmmode~\check{s}\else
  \v{s}\fi{}}}]{Winder20}%
  \BibitemOpen
  \bibfield  {author} {\bibinfo {author} {\bibfnamefont {M.}~\bibnamefont
  {Winder}}, \bibinfo {author} {\bibfnamefont {A.}~\bibnamefont {Hariki}},\
  and\ \bibinfo {author} {\bibfnamefont {J.}~\bibnamefont
  {Kune\ifmmode~\check{s}\else \v{s}\fi{}}},\ }\bibfield  {title} {\bibinfo
  {title} {X-ray spectroscopy of the rare-earth nickelate $\mathrm{LuNiO}_{3}$:
  $\mathrm{LDA}+\mathrm{DMFT}$ study},\ }\href
  {https://doi.org/10.1103/PhysRevB.102.085155} {\bibfield  {journal} {\bibinfo
   {journal} {Phys. Rev. B}\ }\textbf {\bibinfo {volume} {102}},\ \bibinfo
  {pages} {085155} (\bibinfo {year} {2020})}\BibitemShut {NoStop}%
\bibitem [{\citenamefont {Blaha}\ \emph {et~al.}(2018)\citenamefont {Blaha},
  \citenamefont {Schwarz}, \citenamefont {Madsen}, \citenamefont {Kvasnicka},\
  and\ \citenamefont {Luitz}}]{wien2k}%
  \BibitemOpen
  \bibfield  {author} {\bibinfo {author} {\bibfnamefont {P.}~\bibnamefont
  {Blaha}}, \bibinfo {author} {\bibfnamefont {K.}~\bibnamefont {Schwarz}},
  \bibinfo {author} {\bibfnamefont {G.}~\bibnamefont {Madsen}}, \bibinfo
  {author} {\bibfnamefont {D.}~\bibnamefont {Kvasnicka}},\ and\ \bibinfo
  {author} {\bibfnamefont {J.}~\bibnamefont {Luitz}},\ }\href@noop {} {\emph
  {\bibinfo {title} {WIEN2k, An Augmented Plane Wave + Local Orbitals Program
  for Calculating Crystal Properties}}}\ (\bibinfo  {publisher} {Karlheinz
  Schwarz, Techn. Universitat Wien, Austria, 2001, ISBN 3-9501031-1-2},\
  \bibinfo {year} {2018})\BibitemShut {NoStop}%
\bibitem [{\citenamefont {Mostofi}\ \emph {et~al.}(2014)\citenamefont
  {Mostofi}, \citenamefont {Yates}, \citenamefont {Pizzi}, \citenamefont {Lee},
  \citenamefont {Souza}, \citenamefont {Vanderbilt},\ and\ \citenamefont
  {Marzari}}]{wannier90}%
  \BibitemOpen
  \bibfield  {author} {\bibinfo {author} {\bibfnamefont {A.~A.}\ \bibnamefont
  {Mostofi}}, \bibinfo {author} {\bibfnamefont {J.~R.}\ \bibnamefont {Yates}},
  \bibinfo {author} {\bibfnamefont {G.}~\bibnamefont {Pizzi}}, \bibinfo
  {author} {\bibfnamefont {Y.-S.}\ \bibnamefont {Lee}}, \bibinfo {author}
  {\bibfnamefont {I.}~\bibnamefont {Souza}}, \bibinfo {author} {\bibfnamefont
  {D.}~\bibnamefont {Vanderbilt}},\ and\ \bibinfo {author} {\bibfnamefont
  {N.}~\bibnamefont {Marzari}},\ }\bibfield  {title} {\bibinfo {title} {An
  updated version of wannier90: A tool for obtaining maximally-localised
  wannier functions},\ }\href
  {https://doi.org/http://dx.doi.org/10.1016/j.cpc.2014.05.003} {\bibfield
  {journal} {\bibinfo  {journal} {Comput. Phys. Commun.}\ }\textbf {\bibinfo
  {volume} {185}},\ \bibinfo {pages} {2309 } (\bibinfo {year}
  {2014})}\BibitemShut {NoStop}%
\bibitem [{\citenamefont {Kune\v{s}}\ \emph {et~al.}(2010)\citenamefont
  {Kune\v{s}}, \citenamefont {Arita}, \citenamefont {Wissgott}, \citenamefont
  {Toschi}, \citenamefont {Ikeda},\ and\ \citenamefont {Held}}]{wien2wannier}%
  \BibitemOpen
  \bibfield  {author} {\bibinfo {author} {\bibfnamefont {J.}~\bibnamefont
  {Kune\v{s}}}, \bibinfo {author} {\bibfnamefont {R.}~\bibnamefont {Arita}},
  \bibinfo {author} {\bibfnamefont {P.}~\bibnamefont {Wissgott}}, \bibinfo
  {author} {\bibfnamefont {A.}~\bibnamefont {Toschi}}, \bibinfo {author}
  {\bibfnamefont {H.}~\bibnamefont {Ikeda}},\ and\ \bibinfo {author}
  {\bibfnamefont {K.}~\bibnamefont {Held}},\ }\bibfield  {title} {\bibinfo
  {title} {Wien2wannier: {F}rom linearized augmented plane waves to maximally
  localized {W}annier functions},\ }\href
  {https://doi.org/http://dx.doi.org/10.1016/j.cpc.2010.08.005} {\bibfield
  {journal} {\bibinfo  {journal} {Comput. Phys. Commun.}\ }\textbf {\bibinfo
  {volume} {181}},\ \bibinfo {pages} {1888 } (\bibinfo {year}
  {2010})}\BibitemShut {NoStop}%
\bibitem [{\citenamefont {Pavarini}\ \emph {et~al.}(2011)\citenamefont
  {Pavarini}, \citenamefont {Koch}, \citenamefont {Lichtenstein},\ and\
  \citenamefont {Vollhardt}}]{Pavarini1}%
  \BibitemOpen
  \bibfield  {author} {\bibinfo {author} {\bibfnamefont {E.}~\bibnamefont
  {Pavarini}}, \bibinfo {author} {\bibfnamefont {E.}~\bibnamefont {Koch}},
  \bibinfo {author} {\bibfnamefont {A.}~\bibnamefont {Lichtenstein}},\ and\
  \bibinfo {author} {\bibfnamefont {D.~E.}\ \bibnamefont {Vollhardt}},\ }\href
  {http://juser.fz-juelich.de/record/17645} {\emph {\bibinfo {title} {{T}he
  {LDA}+{DMFT} approach to strongly correlated materials}}},\ \bibinfo {series}
  {Schriften des Forschungszentrums J\"ulich : Modeling and Simulation},
  Vol.~\bibinfo {volume} {1}\ (\bibinfo {year} {2011})\ \bibinfo {note} {record
  converted from VDB: 12.11.2012}\BibitemShut {NoStop}%
\bibitem [{\citenamefont {{Pavarini}}(2014)}]{Pavarini2}%
  \BibitemOpen
  \bibfield  {author} {\bibinfo {author} {\bibfnamefont {E.}~\bibnamefont
  {{Pavarini}}},\ }\bibinfo {title} {{Electronic Structure Calculations with
  LDA+DMFT}},\ in\ \href {https://doi.org/10.1007/978-3-319-06379-9_18} {\emph
  {\bibinfo {booktitle} {Many-Electron Approaches in Physics, Chemistry and
  Mathematics, Mathematical Physics Studies, ISBN 978-3-319-06378-2.~Springer
  International Publishing Switzerland}}},\ \bibinfo {editor} {edited by\
  \bibinfo {editor} {\bibfnamefont {V.}~\bibnamefont {{Bach}}}\ and\ \bibinfo
  {editor} {\bibfnamefont {L.}~\bibnamefont {{Delle Site}}}}\ (\bibinfo {year}
  {2014})\ p.\ \bibinfo {pages} {321}\BibitemShut {NoStop}%
\bibitem [{\citenamefont {Boehnke}\ \emph {et~al.}(2011)\citenamefont
  {Boehnke}, \citenamefont {Hafermann}, \citenamefont {Ferrero}, \citenamefont
  {Lechermann},\ and\ \citenamefont {Parcollet}}]{boehnke11}%
  \BibitemOpen
  \bibfield  {author} {\bibinfo {author} {\bibfnamefont {L.}~\bibnamefont
  {Boehnke}}, \bibinfo {author} {\bibfnamefont {H.}~\bibnamefont {Hafermann}},
  \bibinfo {author} {\bibfnamefont {M.}~\bibnamefont {Ferrero}}, \bibinfo
  {author} {\bibfnamefont {F.}~\bibnamefont {Lechermann}},\ and\ \bibinfo
  {author} {\bibfnamefont {O.}~\bibnamefont {Parcollet}},\ }\bibfield  {title}
  {\bibinfo {title} {Orthogonal polynomial representation of imaginary-time
  {G}reen's functions},\ }\href {https://doi.org/10.1103/PhysRevB.84.075145}
  {\bibfield  {journal} {\bibinfo  {journal} {Phys. Rev. B}\ }\textbf {\bibinfo
  {volume} {84}},\ \bibinfo {pages} {075145} (\bibinfo {year}
  {2011})}\BibitemShut {NoStop}%
\bibitem [{\citenamefont {Hafermann}\ \emph {et~al.}(2012)\citenamefont
  {Hafermann}, \citenamefont {Patton},\ and\ \citenamefont
  {Werner}}]{hafermann12}%
  \BibitemOpen
  \bibfield  {author} {\bibinfo {author} {\bibfnamefont {H.}~\bibnamefont
  {Hafermann}}, \bibinfo {author} {\bibfnamefont {K.~R.}\ \bibnamefont
  {Patton}},\ and\ \bibinfo {author} {\bibfnamefont {P.}~\bibnamefont
  {Werner}},\ }\bibfield  {title} {\bibinfo {title} {Improved estimators for
  the self-energy and vertex function in hybridization-expansion
  continuous-time quantum {M}onte {C}arlo simulations},\ }\href
  {https://doi.org/10.1103/PhysRevB.85.205106} {\bibfield  {journal} {\bibinfo
  {journal} {Phys. Rev. B}\ }\textbf {\bibinfo {volume} {85}},\ \bibinfo
  {pages} {205106} (\bibinfo {year} {2012})}\BibitemShut {NoStop}%
\bibitem [{\citenamefont {Hariki}\ \emph {et~al.}(2020)\citenamefont {Hariki},
  \citenamefont {Winder}, \citenamefont {Uozumi},\ and\ \citenamefont
  {Kune\ifmmode~\check{s}\else \v{s}\fi{}}}]{Hariki2020}%
  \BibitemOpen
  \bibfield  {author} {\bibinfo {author} {\bibfnamefont {A.}~\bibnamefont
  {Hariki}}, \bibinfo {author} {\bibfnamefont {M.}~\bibnamefont {Winder}},
  \bibinfo {author} {\bibfnamefont {T.}~\bibnamefont {Uozumi}},\ and\ \bibinfo
  {author} {\bibfnamefont {J.}~\bibnamefont {Kune\ifmmode~\check{s}\else
  \v{s}\fi{}}},\ }\bibfield  {title} {\bibinfo {title}
  {{$\mathrm{LDA}+\mathrm{DMFT}$ approach to resonant inelastic x-ray
  scattering in correlated materials}},\ }\href
  {https://doi.org/10.1103/PhysRevB.101.115130} {\bibfield  {journal} {\bibinfo
   {journal} {Phys. Rev. B}\ }\textbf {\bibinfo {volume} {101}},\ \bibinfo
  {pages} {115130} (\bibinfo {year} {2020})}\BibitemShut {NoStop}%
\bibitem [{\citenamefont {Wang}\ \emph {et~al.}(2009)\citenamefont {Wang},
  \citenamefont {Gull}, \citenamefont {de' Medici}, \citenamefont {Capone},\
  and\ \citenamefont {Millis}}]{wang09}%
  \BibitemOpen
  \bibfield  {author} {\bibinfo {author} {\bibfnamefont {X.}~\bibnamefont
  {Wang}}, \bibinfo {author} {\bibfnamefont {E.}~\bibnamefont {Gull}}, \bibinfo
  {author} {\bibfnamefont {L.}~\bibnamefont {de' Medici}}, \bibinfo {author}
  {\bibfnamefont {M.}~\bibnamefont {Capone}},\ and\ \bibinfo {author}
  {\bibfnamefont {A.~J.}\ \bibnamefont {Millis}},\ }\bibfield  {title}
  {\bibinfo {title} {Antiferromagnetism and the gap of a {M}ott insulator:
  {R}esults from analytic continuation of the self-energy},\ }\href
  {https://doi.org/10.1103/PhysRevB.80.045101} {\bibfield  {journal} {\bibinfo
  {journal} {Phys. Rev. B}\ }\textbf {\bibinfo {volume} {80}},\ \bibinfo
  {pages} {045101} (\bibinfo {year} {2009})}\BibitemShut {NoStop}%
\bibitem [{\citenamefont {Zaanen}\ \emph {et~al.}(1986)\citenamefont {Zaanen},
  \citenamefont {Westra},\ and\ \citenamefont {Sawatzky}}]{Zaanen86}%
  \BibitemOpen
  \bibfield  {author} {\bibinfo {author} {\bibfnamefont {J.}~\bibnamefont
  {Zaanen}}, \bibinfo {author} {\bibfnamefont {C.}~\bibnamefont {Westra}},\
  and\ \bibinfo {author} {\bibfnamefont {G.~A.}\ \bibnamefont {Sawatzky}},\
  }\bibfield  {title} {\bibinfo {title} {Determination of the electronic
  structure of transition-metal compounds: 2$p$ x-ray photoemission
  spectroscopy of the nickel dihalides},\ }\href
  {https://doi.org/10.1103/PhysRevB.33.8060} {\bibfield  {journal} {\bibinfo
  {journal} {Phys. Rev. B}\ }\textbf {\bibinfo {volume} {33}},\ \bibinfo
  {pages} {8060} (\bibinfo {year} {1986})}\BibitemShut {NoStop}%
\bibitem [{\citenamefont {Bocquet}\ \emph {et~al.}(1996)\citenamefont
  {Bocquet}, \citenamefont {Mizokawa}, \citenamefont {Morikawa}, \citenamefont
  {Fujimori}, \citenamefont {Barman}, \citenamefont {Maiti}, \citenamefont
  {Sarma}, \citenamefont {Tokura},\ and\ \citenamefont {Onoda}}]{Bocquet96}%
  \BibitemOpen
  \bibfield  {author} {\bibinfo {author} {\bibfnamefont {A.~E.}\ \bibnamefont
  {Bocquet}}, \bibinfo {author} {\bibfnamefont {T.}~\bibnamefont {Mizokawa}},
  \bibinfo {author} {\bibfnamefont {K.}~\bibnamefont {Morikawa}}, \bibinfo
  {author} {\bibfnamefont {A.}~\bibnamefont {Fujimori}}, \bibinfo {author}
  {\bibfnamefont {S.~R.}\ \bibnamefont {Barman}}, \bibinfo {author}
  {\bibfnamefont {K.}~\bibnamefont {Maiti}}, \bibinfo {author} {\bibfnamefont
  {D.~D.}\ \bibnamefont {Sarma}}, \bibinfo {author} {\bibfnamefont
  {Y.}~\bibnamefont {Tokura}},\ and\ \bibinfo {author} {\bibfnamefont
  {M.}~\bibnamefont {Onoda}},\ }\bibfield  {title} {\bibinfo {title}
  {Electronic structure of early 3$d$-transition-metal oxides by analysis of
  the 2$p$ core-level photoemission spectra},\ }\href
  {https://doi.org/10.1103/PhysRevB.53.1161} {\bibfield  {journal} {\bibinfo
  {journal} {Phys. Rev. B}\ }\textbf {\bibinfo {volume} {53}},\ \bibinfo
  {pages} {1161} (\bibinfo {year} {1996})}\BibitemShut {NoStop}%
\bibitem [{\citenamefont {Bocquet}\ \emph {et~al.}(1992)\citenamefont
  {Bocquet}, \citenamefont {Mizokawa}, \citenamefont {Saitoh}, \citenamefont
  {Namatame},\ and\ \citenamefont {Fujimori}}]{Bocquet92}%
  \BibitemOpen
  \bibfield  {author} {\bibinfo {author} {\bibfnamefont {A.~E.}\ \bibnamefont
  {Bocquet}}, \bibinfo {author} {\bibfnamefont {T.}~\bibnamefont {Mizokawa}},
  \bibinfo {author} {\bibfnamefont {T.}~\bibnamefont {Saitoh}}, \bibinfo
  {author} {\bibfnamefont {H.}~\bibnamefont {Namatame}},\ and\ \bibinfo
  {author} {\bibfnamefont {A.}~\bibnamefont {Fujimori}},\ }\bibfield  {title}
  {\bibinfo {title} {Electronic structure of 3$d$-transition-metal compounds by
  analysis of the 2$p$ core-level photoemission spectra},\ }\href
  {https://doi.org/10.1103/PhysRevB.46.3771} {\bibfield  {journal} {\bibinfo
  {journal} {Phys. Rev. B}\ }\textbf {\bibinfo {volume} {46}},\ \bibinfo
  {pages} {3771} (\bibinfo {year} {1992})}\BibitemShut {NoStop}%
\bibitem [{\citenamefont {Matsubara}\ \emph {et~al.}(2005)\citenamefont
  {Matsubara}, \citenamefont {Uozumi}, \citenamefont {Kotani},\ and\
  \citenamefont {Claude~Parlebas}}]{Matsubara05}%
  \BibitemOpen
  \bibfield  {author} {\bibinfo {author} {\bibfnamefont {M.}~\bibnamefont
  {Matsubara}}, \bibinfo {author} {\bibfnamefont {T.}~\bibnamefont {Uozumi}},
  \bibinfo {author} {\bibfnamefont {A.}~\bibnamefont {Kotani}},\ and\ \bibinfo
  {author} {\bibfnamefont {J.}~\bibnamefont {Claude~Parlebas}},\ }\bibfield
  {title} {\bibinfo {title} {Charge {T}ransfer {E}xcitation in {R}esonant
  {X}-ray {E}mission {S}pectroscopy of {N}i{O}},\ }\href
  {https://doi.org/10.1143/JPSJ.74.2052} {\bibfield  {journal} {\bibinfo
  {journal} {J. Phys. Soc. Jpn.}\ }\textbf {\bibinfo {volume} {74}},\ \bibinfo
  {pages} {2052} (\bibinfo {year} {2005})}\BibitemShut {NoStop}%
\bibitem [{\citenamefont {Park}\ \emph {et~al.}(1988)\citenamefont {Park},
  \citenamefont {Ryu}, \citenamefont {Han},\ and\ \citenamefont {Oh}}]{Park88}%
  \BibitemOpen
  \bibfield  {author} {\bibinfo {author} {\bibfnamefont {J.}~\bibnamefont
  {Park}}, \bibinfo {author} {\bibfnamefont {S.}~\bibnamefont {Ryu}}, \bibinfo
  {author} {\bibfnamefont {M.-S.}\ \bibnamefont {Han}},\ and\ \bibinfo {author}
  {\bibfnamefont {S.-J.}\ \bibnamefont {Oh}},\ }\bibfield  {title} {\bibinfo
  {title} {Charge-transfer satellites in the 2$p$ core-level photoelectron
  spectra of heavy-transition-metal dihalides},\ }\href
  {https://doi.org/10.1103/PhysRevB.37.10867} {\bibfield  {journal} {\bibinfo
  {journal} {Phys. Rev. B}\ }\textbf {\bibinfo {volume} {37}},\ \bibinfo
  {pages} {10867} (\bibinfo {year} {1988})}\BibitemShut {NoStop}%
\end{thebibliography}%


\end{document}